# On Achieving Local View Capacity Via Maximal Independent Graph Scheduling

Vaneet Aggarwal, A. Salman Avestimehr, and Ashutosh Sabharwal

*Abstract*—"If we know more, we can achieve more." This adage also applies to communication networks, where more information about the network state translates into higher sum-rates. In this paper, we formalize this increase of sum-rate with increased knowledge of the network state. The knowledge of network state is measured in terms of the number of hops, $h$, of information available to each transmitter and is labeled as $h$-local view. To understand how much capacity is lost due to limited information, we propose to use the metric of normalized sum-capacity, which is the $h$-local view sum-capacity divided by global-view sum capacity. For the cases of one and two-local view, we characterize the normalized sum-capacity for many classes of deterministic and Gaussian interference networks. In many cases, a scheduling scheme called maximal independent graph scheduling is shown to achieve normalized sum-capacity. We also show that its generalization for 1-local view, labeled coded set scheduling, achieves normalized sum-capacity in some cases where its uncoded counterpart fails to do so.

## I. INTRODUCTION

### A. Overview

Node mobility in wireless networks leads to constant changes in network connectivity at long time-scales and per link channel gains at short time-scales. The optimal rate allocation and associated encoding and decoding rules depend on both the network connectivity and the current channel gains of all links (commonly referred as network state). However, in large wireless networks, acquiring full network connectivity and state information for making optimal decisions is typically infeasible. Thus, in the absence of centralization of network state information, nodes have limited local view of the whole network. As a result, the local view of the nodes are mismatched and different from local views of other nodes. Thus, each node has potentially a different snapshot of the whole network. Due to mismatched local views, nodes' decisions about their transmission (like rate, power, codebook) and reception (method of decoding) parameters are inherently distributed. The key question then is how do optimal distributed decisions

V. Aggarwal is with AT&T Shannon Labs, 180 Park Ave - Building 103, Florham Park NJ 07932, USA (email: vaneet@research.att.com). He was with Department of Electrical Engineering, Princeton University, Princeton NJ 08544, when this work was done. A. S. Avestimehr is with School of Electrical and Computer Engineering, Cornell University, Ithaca NY, USA (email: avestimehr@ece.cornell.edu). The research of A. S. Avestimehr was supported in part by the NSF CAREER award 0953117. A. Sabharwal is with Department of Electrical and Computer Engineering, Rice University, Houston TX 77005, USA (email: ashu@rice.edu). The paper was presented in part at the Allerton Conference on Communication, Control and Computing 2009 [9], the IEEE Asilomar Conference on Signals, Systems and Computers 2009, the IEEE Conference on Information Sciences and Systems 2010 and the IEEE International Symposium on Information Theory 2010.

perform when compared to the optimal decisions which have full network state information.

We immediately acknowledge the difficulty in answering the above question. Even with full global information, where each node knows the full network connectivity and current state perfectly, the capacity of general networks is an open problem. In light of that fact, our driving question adds additional complexity to the analysis by asking nodes to rely only on their local views. To make progress, we make several simplifying assumptions in our choice of network model and the model for local view. Even in the simplified model, our analysis leads to several significant conclusions as described below.

In this paper, we limit our attention to $K$-user single-hop interference networks with $K$ transmitters and $K$ receivers. Each transmitter communicates with its receiver in a single-hop fashion but in the process can interfere with an arbitrary number of receivers. The special cases include the classic two-user interference network, $Z$-network, one-to-many, many-to-one and fully-connected interference networks. In this paper, we will consider both the linear deterministic [10, 11] and the Gaussian models for the network.

To model the local view, we will borrow the concept of hop distance from networking literature and consider the case where each transmitter has a perfect knowledge of all links within $h$ hops from it and has no knowledge of links beyond $h$ hops. As a result, if $h$ is less than the network diameter, a subset of transmitters will end with mismatched knowledge about the state of the channels. Since each channel gain can range from zero to a maximum value, our formulation is similar to compound channels [1, 2] with one major difference. In the multi-terminal compound network formulations, *all* nodes are missing identical information about the channels in the network. In our formulation, the hop-based model of local view leads to nodes with *asymmetric* information about the channels in the network. Thus to emphasize that the lack of knowledge is asymmetric, we have labeled the resulting compound channel capacity formulation as *local view capacity*. Finally, we assume that the nodes know the connectivity, i.e., which pairs of the links can exist but may or may not know the actual value of the channel gains on those links. In graph-theoretic parlance, the nodes are assumed to know the edges of the graph (i.e. the shape of the network) but not their weights which represent channel gains. This is partially motivated by the fact that the network connectivity often changes at a much slower time-scales than the channel gains.

Finally, realizing the difficulty of directly characterizing capacity (sum or the whole region), we propose to study the best guaranteed ratio of the sum rate with local view to the sum-capacity with full global view at each node. We label this as normalized sum-capacity, $\alpha^* \in [0, 1]$. As shown in Figure 1, our goal is to characterize the normalized sum-capacity as a function of the hops of information about the network that is available at the nodes. In many cases, it turns out that the normalized capacity is easier to characterize than the actual capacity since this involves finding sum-capacity for a smaller range of the values of channel gains.

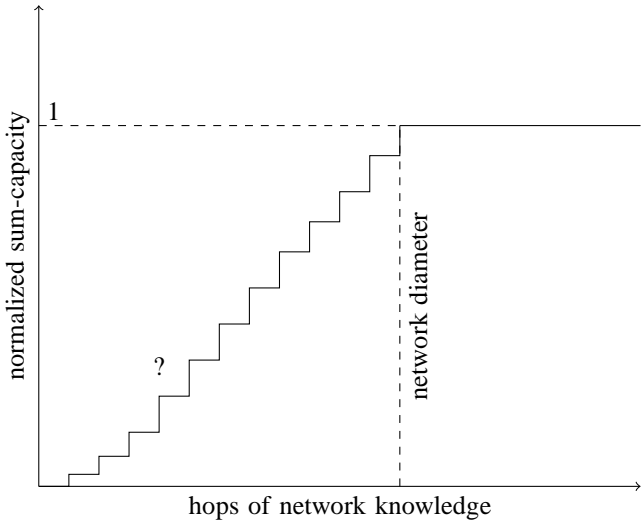

Fig. 1. Increase of normalized sum-capacity with the hops of information about the network.

*B. Main Contributions*

Our objective is to maximize global sum-rate with mismatched local views. However, nodes have to base their decision only on their local asymmetric views which in turn implies that their decisions are naturally distributed. One intuitive solution is for nodes to coordinate their transmissions such that the nodes beyond $h$ hops transmit only if they can cause no interference with $h$-hop size sub-network and thus each connected sub-network operates as if it is a network with full global information. This is formalized through the notion of an independent graph, which is defined as a sub-graph which admits a distributed encoding and decoding scheme which achieves same sum-capacity as a scheme with full global information. We use this intuition to propose *maximal independent graph scheduling*, where the network is divided into sub-graphs (equivalently sub-networks) and the sub-graphs are scheduled orthogonally over time. The sub-graphs are chosen such that they are maximal independent graphs which ensure highest spatial reuse of the users.

For one hop information at the transmitters, maximal independent graphs are equivalent to maximal independent sets (MIS), which are largest subsets with non-interfering transmitter-receiver pairs. Note that maximal independent set scheduling or maximal weighted independent sets are often the optimal schedules under traditional SINR (Signal to Interference plus Noise Ratio) based protocol models for networks [3]. Our results show that the MIS schedule is information-theoretically optimal in several cases. Hence, we provide an information-theoretic notion of optimality for the MIS scheduling algorithm in those cases.

We show that in several cases, a maximal independent graph (MIG) scheduling algorithm achieves the maximum normalized sum-rate among all distributed encoding and decoding schemes, when the transmitters have no more than two hops of channel information. The MIG schedule is shown to be optimal for most three-user bipartite interference topologies, $K$-user cyclic chain, $K$-user $d$-to-many interference network, etc.

However, we show that the MIG schedule is not optimal in general for all network topologies and higher rates can be achieved by exploiting coding. For example, in the case of 1-local view in 3-user cyclic chain network, we show that a coded set (CS) schedule, where the coding is performed over two scheduling time-slots, achieves a higher normalized sum-rate than pure scheduling. In CS scheduling, receivers of inactive transmitters continue listening and train themselves on the interference caused by other nodes. Then, they use this interference in a later slot to aid reliable decoding of their own codeword. For linear deterministic interference networks of [10] with 1-local view, we also give an algorithm that achieves normalized sum-capacity.

*C. Related Work*

The work on understanding role of limited network knowledge was first initiated in [6, 7], where the authors used a message-passing abstraction of network protocols to formulate a metric of limited network view at each node in the form of number of message rounds; each message round adds two extra hops of channel information at the transmitters. The key result was that distributed decisions can be either sum-rate optimal or can be arbitrarily worse than the global-information sum-capacity. This result was further strengthened for arbitrary $K$-user interference network in [9], where the authors characterized all network connectivities to allow optimal distributed rate allocation with two hops of network information at each transmitter. In this paper, we take the next major step in understanding the performance of distributed decisions. We compute the capacity of distributed decisions for several network topologies with one-hop and two-hop network information at the transmitter.

The rest of the paper is organized as follows. In Section II, we give the system and network model, and provide some definitions that will be used throughout the paper. We will also consider an example of Multiple Access Network to gain understanding. In Section III, we define maximal independent graph scheduling and derive the independent graphs in the cases when the transmitters have 1 or 2 hops channel gain information. In Section IV, we characterize the cases where maximal independent graph scheduling is optimal. In Section V, we give example where maximal independent graph scheduling is not optimal, and extend the achievable scheme with 1-hop knowledge at transmitters to coded set scheduling. We also give the optimal algorithm with 1-hop knowledge at the transmitter for the linear deterministic model of [10].

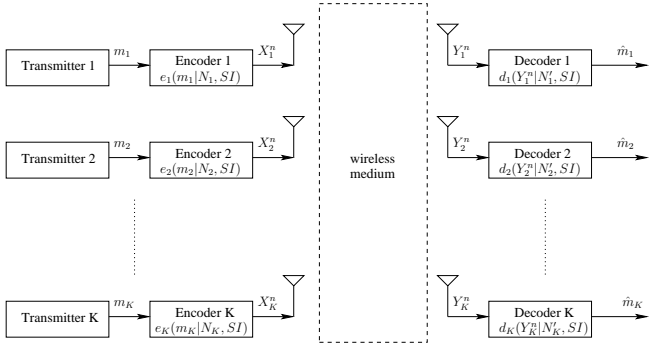

Fig. 2. System-level depiction of the problem.

Section VI considers 3 hops of knowledge at the transmitters and Section VII concludes the paper.

## II. PROBLEM FORMULATION

In this section, we will first describe the system and network models. We will then define normalized sum-rate and normalized sum-capacity which will be used to evaluate the performance with asymmetric network information at the nodes. Finally, we will also formalize the specific notion of local view used in this paper to model asymmetric network information.

### A. System model

As shown in Figure 2, consider a wireless network with $K$ transmitters and $K$ receivers. Each node in the network is either a transmitter or a receiver. For each transmitter $k$, let the message index $m_k$ be encoded as $X_k^n$ using the encoding functions $e_k(m_k|N_k, \mathsf{SI})$, which depend on the local view, $N_k$, and side information about the network, $\mathsf{SI}$. Only receiver $k$ is interested in message $m_k$. The message is decoded at the receiver $k$ using the decoding function $d_k(Y_k^n|N_k', \mathsf{SI})$, where $N_k'$ is the receiver local view and $\mathsf{SI}$ is the side information. A strategy is defined as the set of all encoding and decoding functions in the network, $\{e_k(m_k|N_k, \mathsf{SI}), d_k(Y_k^n|N_k', \mathsf{SI})\}$. We note that the local view at transmitter $k$ and receiver $k$ can be different, as will be the case in our subsequent development. The relationship between the transmit signals and the received signals is specified by the network model that is described in the next section.

### B. Network Model

We will consider two models for interference networks. We use a deterministic model, which was proposed as an approximation to the Gaussian model in [10] to get insights and then proceed to Gaussian network model both of which are described as follows.

*1) Linear Deterministic Model [10]:* In a linear deterministic interference network, the input of the $k^{\text{th}}$ transmitter at time $i$ can be written as $X_k[i] = \begin{bmatrix} X_{k_1}[i] & X_{k_2}[i] & \ldots X_{k_q}[i] \end{bmatrix}^T$, $k = 1, 2, \cdots, K$, such that $X_{k_1}[i]$ and $X_{k_q}[i]$ are the most and the least significant bits, respectively. The received signal of user $j$, $j = 1, 2, \cdots, K$, at time $i$ is denoted by the vector $Y_j[i] = \begin{bmatrix} Y_{j_1}[i] & Y_{j_2}[i] & \ldots & Y_{j_q}[i] \end{bmatrix}^T$. Associated with each transmitter $k$ and receiver $j$ is a non-negative integer $n_{kj}$ that represents the gain of the channel between them. The maximum number of bits supported by any link is $q = \max_{k,j}(n_{kj})$. The received signal $Y_j[i]$ is given by

$$Y_j[i] = \sum_{k=1}^{K} \mathbf{S}^{q-n_{kj}} X_k[i], \quad (1)$$

where $q$ is the maximum of the channel gains (*i.e.* $q = \max_{j,k}(n_{jk})$), the summation is in $\mathbb{F}_2^q$, and $\mathbf{S}^{q-n_{jk}}$ is a $q \times q$ shift matrix with entries $\mathbf{S}_{m,n}$ that are non-zero only for $(m,n) = (q - n_{jk} + n, n), n = 1, 2, \ldots, n_{jk}$. We will also use $X_k^n$, $Y_k^n$ to denote $(X_{k1}, \cdots, X_{kn})$, $(Y_{k1}, \cdots, Y_{kn})$. The network can be represented by a square matrix $H$ whose $(i,j)^{th}$ entry is $H_{ij} = n_{ij}$. We note that $H$ need not be symmetric.

*2) Gaussian Model:* In a Gaussian interference network, the inputs of the $k^{\text{th}}$ transmitter at time $i$ are denoted by $X_k[i] \in \mathbb{C}$, $k = 1, 2, \cdots, K$, and the outputs at $j^{\text{th}}$ receiver in time $i$ can be written as $Y_j[i] \in \mathbb{C}$, $j = 1, 2, \cdots, K$. The received signal $Y_j[i]$, $j = 1, 2, \cdots, K$ is given by

$$Y_j[i] = \sum_{k=1}^{K} h_{kj} X_k[i] + Z_j[i], \quad (2)$$

where $h_{kj} \in \mathbb{C}$ is the channel gain associated with each transmitter $k$ and receiver $j$, and $Z_j[i]$ are additive white complex Gaussian random variables of unit variance. Much like the deterministic case, we will use $X_k^n$, $Y_k^n$ to denote $(X_k[1], \cdots, X_k[n])$, $(Y_k[1], \cdots, Y_k[n])$. Further, the input $X_k[i]$ has an average power constraint of unity, i.e. $\mathbb{E}(\frac{1}{n}\sum_{i=1}^{n}|X_k[i]|^2) \leq 1$, where $\mathbb{E}$ denotes the expectation of the random variable.

Like the deterministic case, we represent the network by a square matrix $H$ whose $(i,j)^{th}$ entry is $H_{ij} = |h_{ij}|^2$ and can similarly define the set of network states. Thus we will use the matrix $H$ for both the deterministic and the Gaussian model, where the usage will be clear from the context.

### C. Normalized sum-capacity

As we discussed earlier, at each receiver $k$, the desired message $m_k$ is decoded using the decoding function $d_k(Y_k^n|N_k', \mathsf{SI})$, where $N_k'$ is the receiver local view of the network and $\mathsf{SI}$ is the side information. The corresponding probability of decoding error $\lambda_j(n)$ is defined as $\Pr[m_k \neq d_k(Y_k^n|N_k', \mathsf{SI})]$. A rate tuple $(R_1, R_2, \cdots, R_K)$ is said to be achievable if there exists a sequence of codes such that the error probabilities $\lambda_1(n), \cdots \lambda_K(n)$ go to zero as $n$ goes to infinity for all network states consistent with the side information. The sum-capacity is the supremum of $\sum_i R_i$ over all possible encoding and decoding functions.

We will now define normalized sum-rate and normalized sum-capacity that will be used throughout the paper. These notions represent the percentage of the global-view sum-capacity that can be achieved with partial information about the network.

**Definition 1.** *Normalized sum-rate of $\alpha$ is said to be achievable for a set of network states with partial information if there exists a strategy such that following holds. The strategy yields a sequence of codes having rates $R_i$ at the transmitter $i$ such that the error probabilities at the receiver, $\lambda_1(n), \cdots \lambda_K(n)$, go to zero as $n$ goes to infinity, satisfying*

$$\sum_i R_i \geq \alpha C_{sum} - \tau$$

*for all the sets of network states consistent with the side information, and for a constant $\tau$ that is independent of the channel gains but may depend on the side information* SI. *Here $C_{sum}$ is the sum-capacity of the whole network with the full information.*

**Definition 2.** *Normalized sum-capacity, $\alpha^*$, is defined as the supremum over all achievable normalized sum rates $\alpha$.*

Note that $\alpha^* \in [0, 1]$. In [7], we defined the concept of *universal optimality* of a strategy. A universally optimal strategy is the one which achieves $\alpha^*(h) = 1$ for a given network. Thus, universal optimality is the special case where the distributed scheme achieves global-view sum-capacity in *all* network states and hence is universally optimal for all network states.

*D. Local View Based on Hop Distance*

We assume that that there is a direct link between each transmitter $T_i$ and its intended receiver $D_i$. On the other hand, if a cross-link between transmitter $i$ and receiver $j$ does not exist, then $H_{ij} \equiv 0$. For large part, we will treat the network as a weighted undirected graph, $G = (V, E, W)$, where transmitters and receivers are the vertices of the graph, $V = \{T_i, D_i\}$, and an edge $e \in E$ exists between any two nodes if they have a possibility of non-zero channel gain. In other words, if the channel gain between two nodes is identically zero, there is no edge between them[1]. Finally, the actual channel gain $n_{ij}$ (for deterministic model) or $h_{ij}$ (for Gaussian model) is the edge weight $w(e) \in W$. The resulting bipartite graph thus has $2K$ vertices and no more than $K^2$ edges.

We realize that the current formulation of distributed encoding is very general and encompasses a large class of $\{N_k, N'_k\}_k$ and SI. To make progress we will focus on a special structure of local view and side information at the nodes, which is largely inspired by common characteristics of existing network protocols. We will assume that the side information at all the nodes is the network connectivity characterized by $(E, V)$. We identify $(E, V)$ with the long time-scale characteristics of the network, which changes slowly. However, the network state captured by edge weights $W$, which gives the weights of edges is not part in the side information.

The local view at the nodes is defined using the metric of hop count ($h$). For any node, the links that are incident on the node have a distance of 1-hop. In general, hop-distance of a link from a node is one plus the minimum amount of links traversed starting from the node and terminating at the link. An example of the minimum distance of the links from a node is shown in Figure 3. We say that there is $h$-local view when all the transmitters know the weights (equivalently the channel gains) of those links which are at a distance of $h$-hops from them while the receivers know the weight of only those links which are at most distance of $h + 1$ hops from them. This definition of $h$-local information is based on our prior work in [7] where we proposed a multi-round protocol abstraction to show how different nodes have different amounts of network information. In the message-passing abstraction, it was convenient to have receivers know one more hop than their corresponding transmitters, which allowed coherent decoding.

[1] The model is inspired by fading channels, where the existence of a link is based on its average channel gain. On the average the link gain may be above the noise floor but its instantaneous value can be below the noise floor.

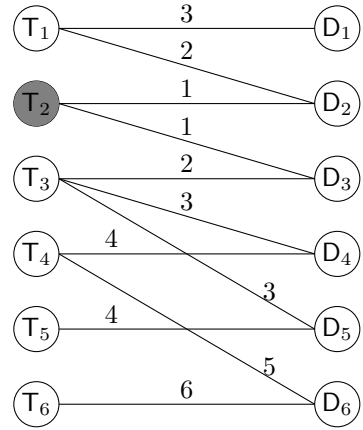

Fig. 3. The hop-distances of each link from transmitter, $T_2$ (the dark circle), are labeled above each link.

Thus, we will consider the side information SI to be the network connectivity while the local information at each node is the $h$-local information. Thus, each transmitter uses a codebook of rate $R_i$ which is a function of network connectivity and local channel gain information. A strategy at the transmitters achieves normalized sum-rate of $\alpha$ if the sum rate achieved is within a constant bits of $\alpha$ times the sum capacity with global knowledge of all the channel gains in the network for all sets of channel gains possible in the network. As $h$ increases, the normalized sum-capacity increases. When $h$ is the network diameter, which is the maximum hop distance between any link and any node, all the nodes have full network information. This is called the global view, since every node knows the complete network state, $G = (V, E, W)$. In this setting, normalized sum-capacity $\alpha^* = 1$. When $h = 0$, none of the nodes know any weights and thus following compound channel arguments [1], $\alpha^* = 0$ since none of the nodes know any link weight and have to assume that all channel gains are zero.

*E. A Warmup Example: Multiple Access Network*

We start with a simple example to illustrate these concepts. As shown in Figure 4, we consider the $K$-user Gaussian multiple access network with the channel gain from $i^{th}$ transmitter to the receiver being $h_i$ such that $|h_i|^2 = \sqrt{\mathsf{SNR}_i}$ and the power constraint at each transmitter being unity. Note that the network diameter is two, which implies 2-local is equivalent to global view implying $\alpha^*(2) = 1$. Thus the interesting case is that of 1-local view.

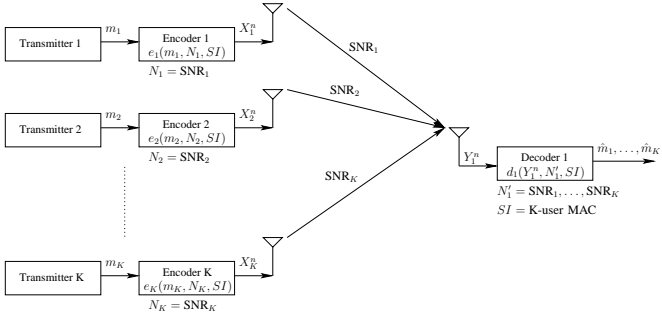

Fig. 4. Example: multiple-access network with 1-hop local information.

We show that when there is 1-local view, the normalized sum-capacity is $1/K$ which can be achieved by simply scheduling one user at a time in a total of $K$ time-slots. It can also be achieved by letting each user simultaneously send at $1/K$ fraction of its direct link capacity.

The main challenge is to show the converse. Let $K > 1$, as otherwise the result holds trivially. Assume that normalized sum-rate of $\alpha = (1/K + \epsilon)$ is achievable. Then, we should be able to achieve a rate tuple satisfying

$$R_i \geq \left(\frac{1}{K} + \epsilon\right) \log(1 + \mathsf{SNR}_i) - \tau, \ \forall \ 1 \leq i \leq K. \quad (3)$$

This is because each node is unaware of the other channel gains. To achieve a normalized sum-rate larger than $\alpha$, each user should send at a rate larger than a fraction $\alpha$ of its channel capacity up to a difference $\tau$ (otherwise in the case when all other channel gains are zero, achievable normalized sum-rate is smaller than $\alpha$). Now, we will show that this rate-tuple cannot be achieved. With the capacity bound of full information,

$$\begin{aligned}
R_K &\leq \log\left(1 + \sum_{i=1}^{K} \mathsf{SNR}_i\right) - \sum_{i=1}^{K-1} R_i \\
&\stackrel{(3)}{\leq} \log\left(1 + \sum_{i=1}^{K} \mathsf{SNR}_i\right) \\
&\quad - \left(\frac{1}{K} + \epsilon\right) \sum_{i=1}^{K-1} \log(1 + \mathsf{SNR}_i) + (K-1)\tau.
\end{aligned}$$

Since the $K^{th}$ transmitter does not know $\mathsf{SNR}_i$ for $1 \leq i \leq K - 1$,

$$\begin{aligned}
R_K &< \min_{\mathsf{SNR}_i, 1 \leq i \leq K-1} \left[\log\left(1 + \sum_{i=1}^{K} \mathsf{SNR}_i\right)\right. \\
&\qquad \left. - \left(\frac{1}{K} + \epsilon\right) \sum_{i=1}^{K-1} \log(1 + \mathsf{SNR}_i) + (K-1)\tau\right] \\
&\leq \frac{1}{K} \log(1 + \mathsf{SNR}_K) \\
&\quad -(K-1)\epsilon \log(1 + \mathsf{SNR}_K) + \log(K) \\
&\quad +(K-1)\tau \quad (4)
\end{aligned}$$

For the above to hold, $(K-1)\epsilon \log(1+\mathsf{SNR}_K) \leq \log(K) + (K-1)\tau$ which cannot hold for all $\mathsf{SNR}_K$ with $\tau$ and $K$ independent of $\mathsf{SNR}_K$. Thus, $\alpha^* \leq \frac{1}{K}$.

Since all the links are at-most two hops from each transmitter, the normalized sum-capacity in the case when each transmitter knows all the links that are at-most two hop distant from it is 1.

For the rest of the paper, we will focus on interference networks some examples of which will be defined in the next section.

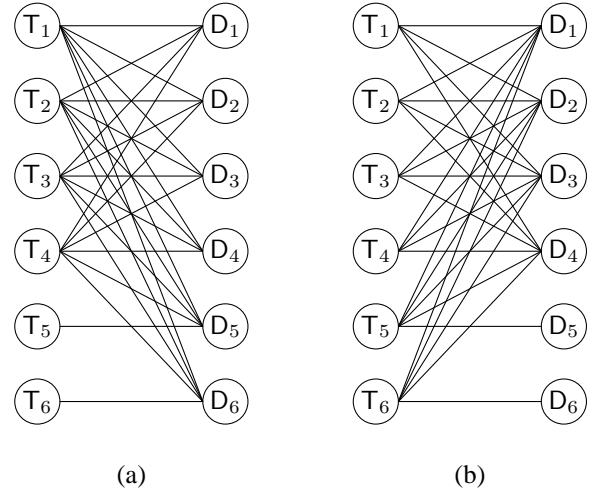

Fig. 5. (a) 4-to-many interference network, and (b) many-to-4 interference network with 6 users.

### F. Examples of Interference Networks

In this paper, some special interference networks will be used as examples. They are defined as follows.

**Definition 3.** *A <u>d-to-many</u> interference network with $K$ users is an interference network specified by* $\mathsf{E} = \bigcup_{i=1}^{K}\{(\mathsf{T}_i, \mathsf{D}_i)\} \bigcup \bigcup_{i=1}^{d} \bigcup_{j=1}^{K}\{(\mathsf{T}_i, \mathsf{D}_j)\}$. *This network has links from the first $d$ transmitters to all the receivers.*

**Definition 4.** *A <u>many-to-$d$</u> interference network of $K$ users is an interference network specified by* $\mathsf{E} = \bigcup_{i=1}^{K}\{(\mathsf{T}_i, \mathsf{D}_i)\} \bigcup \bigcup_{i=1}^{K} \bigcup_{j=1}^{d}\{(\mathsf{T}_i, \mathsf{D}_j)\}$. *This network has links from all transmitters to the first $d$ receivers.*

Example of 4-to-many interference network and many-to-4 interference networks with 6 users are depicted in Figure 5.

**Definition 5.** *A <u>fully-connected</u> interference network with $K$ users is many-to-$K$ interference network with $K$ users which is also the same as a $K$-to-many interference network with $K$ users.*

**Definition 6.** *A <u>chain</u> of $K$ users is an interference network defined by* $\mathsf{E} = \bigcup_{i=1}^{K}\{(\mathsf{T}_i, \mathsf{D}_i)\}$
$\bigcup \bigcup_{i=1}^{K-1}\{(\mathsf{T}_i, \mathsf{D}_{i+1})\}$. *This network has links from each transmitter to its next receiver. A Z-network is a chain of 2 users.*

**Definition 7.** *A <u>cyclic-chain</u> of $K$ users is an interference network defined by* $\mathsf{E} = \bigcup_{i=1}^{K}\{$

$(\mathsf{T}_i, \mathsf{D}_i)\} \bigcup \bigcup_{i=1}^{K-1}\{(\mathsf{T}_i, \mathsf{D}_{i+1})\} \bigcup \{(\mathsf{T}_K, \mathsf{D}_1)\}$. *This network is similar to a $K-user$ chain of Definition 6 except that the last transmitter interferes with the first receiver, thereby making the network a circular chain.*

## III. Subgraph Scheduling

In this section, we will present a scheduling-based scheme which uses partial information at every node. The main idea is to divide the network into smaller disjoint sub-networks, each of which can operate optimally such that the normalized sum-rate of $\alpha^*(h) = 1$ for each sub-network. The choice of sub-networks thus becomes important and will be addressed in the form of independent sub-graphs as discussed below.

We will use the graph-theoretic terminology introduced in Section II-D to describe the scheduling algorithm. The graph theoretic formulation will allow us to compare our results to existing results in the literature for the special case of single-hop local view, as discussed in Section IV. Further, the graph-theoretic formulation will facilitate parallels between our proposed scheduling method and graph-concepts of chromatic number, again discussed in Section IV.

In Section III-A, we will first describe the scheduling algorithm and derive its achievable normalized sum-rate performance for arbitrary hop-view, assuming independent graphs are known. In Section III-B, we will derive the form of independent sub-graphs for 1- and 2-local view. An example is provided in Section III-C.

### A. Maximal Independent Graph Scheduling

Following standard graph theory terminology, a subgraph $A \subseteq \mathsf{G}$, is a subset of vertices and edges in $\mathsf{G}$. The complement of $A$ is $A^c$ such that $(\mathsf{V}, \mathsf{E}) = A \cup A^c$. In this section, we will only consider subgraphs where both transmitter $\mathsf{T}_i$ and its corresponding receiver $\mathsf{D}_i$ are either in the subgraph together or in its complement. We will remove this restriction on subgraphs in Section V to propose a generalization which can achieve strictly higher rates for some networks compared to the following sub-graph schedule. Note that while the graph edges are weighted with the channel gains, the edge weights will not play a role in the description of the scheduling algorithm. Hence in our definition of subgraphs, we do not include edge weights. Since the network connectivity is known as side information to all the nodes and the schedules only depend on the connectivity, each user knows the schedule and hence when to transmit or when not to transmit.

With the above (restricted) definition of subgraph, any strict subgraph $A \subseteq \mathsf{G}$ represents a valid interference network with a reduced number of transmitter-receiver pairs. For that subgraph $A$, the normalized sum-rate $\alpha_A^*(h)$ can be defined, which is the ratio of sum-capacity with $h$-local view to the sum-capacity with global view ($h = \text{diameter}(A)$) for network $A$.

Armed with the above framework, we can now define Independent Graph Scheduling as follows. Let $A_1, A_2, \ldots, A_t$ be $t$ sub-graphs (not necessarily distinct) of the network $\mathsf{G}$ such that for each sub-graph $A_i$, $\alpha_{A_i}^*(h) = 1$. Subgraphs for which $\alpha_{A_i}^*(h) = 1$ are called independent subgraphs. Since transmitter-receiver pairs are either part of $A_i$ or $A_i^c$, each pair either appears in a subgraph $A_i$ or it does not appear in $A_i$.

**Definition 8** (Independent Graph Scheduling). *Independent Graph Scheduling parametrized by $t$ independent sub-graphs $A_1, A_2, \ldots, A_t$ uses $t$ time-slots and schedules the sub-graph $A_i$ in time-slot $i$.*

Define the indicator function

$$\mathbf{1}_{j \in A_i} = \begin{cases} 1 & \mathsf{T}_j \in A_i \\ 0 & \mathsf{T}_j \notin A_i \end{cases}. \quad (5)$$

For *any* given tuple of independent subgraphs, $\{A_i\}_{i=1}^t$, which satisfy $\alpha_{A_i}^*(h) = 1$, the next theorem gives the normalized sum-rate that can be achieved by sub-graph scheduling.

**Theorem 1** (Achievable Normalized Sum-rate of Independent Graph Scheduling). *Independent Graph Scheduling parametrized by $t$ independent sub-graphs $A_1, A_2, \ldots, A_t$ achieves a normalized sum-rate of $d/t$, where*

$$d = \min_{j \in \{1, 2, \ldots, K\}} \sum_{i=1}^t \mathbf{1}_{j \in A_i}. \quad (6)$$

*Proof:* Let $(C_1, \cdots, C_K)$ be any point in the full knowledge capacity region. The achievable rate in time-slot $i$ is $R^{(i)} \geq \sum_{\{j\} \subseteq A_i} C_i - \tau_i$ by the choice of subgraphs $A_i$ which satisfy $\alpha_{A_i}^*(h) = 1$. Note that $\tau$ is dependent on $i$ since it can change in each time-slot due to selection of different subgraphs. Hence, the overall rate is $\frac{1}{t}\sum_{i=1}^t R_i \geq \frac{1}{t}\sum_{i=1}^t \sum_{\{j\} \subseteq A_i} C_i - \frac{1}{t}\sum_{i=1}^t \tau_i \geq \frac{d}{t}(C_1 + \cdots + C_K) - \frac{1}{t}\sum_{i=1}^t \tau_i$. By the definition of normalized sum-rate, $\alpha = d/t$. ∎

First note that the sub-graphs $A_i$ need not be distinct, which allows allocating more than one time-slot to a particular subgraph if needed. Second, the subgraph set $\{A_i\}_{i=1}^t$ and the number of subgraphs $t$ are both design variables and should be chosen to maximize $d/t$, such that the overall network rate is maximized. The $d/t$-maximizing choice of subgraphs is labeled as a *maximal independent graph* (MIG) schedule.

The main idea behind MIG scheduling is to decouple transmissions of nodes from the unknown part of the network. This is done by switching off some of the flows such that the network gets partitioned into disconnected subgraphs. However, switching off flows means potentially lost rate compared to global-view optimal sum-capacity, so the subgraphs have to be selected to maximize *spatial reuse*. That is, this involves operating as many flows as possible in parallel while still satisfying $\alpha_{A_i}^*(h) = 1$. Such subgraphs are labeled *maximal independent graphs* and form the core of MIGS. We characterize independent graphs next.

### B. Identifying Independent Graphs

Since MIG scheduling schedules a subgraph $A_i$ satisfying $\alpha_{A_i}^*(h) = 1$ in time-slot $i$, we need a characterization of independent sub-graphs. The problem turns out to be very challenging for a general $h$. We provide complete characterization for two important cases of $h = 1$ and $h = 2$, for both deterministic and Gaussian networks, in the next two theorems. The special case of $h = 2$ for the deterministic networks was presented in [9]; in this paper, we provide a tight outer bound and also extend it to Gaussian networks.

We note that the sufficient and necessary conditions in following two theorems are stated in terms of the graph properties of G. Theorem 2 uses the node degree, which is the number of edges incident on the node. Theorem 3 uses the definitions in Section II-F.

**Theorem 2** (1-local View Independent Subgraphs). *The normalized sum-capacity of a $K$-user interference network (deterministic or Gaussian) with 1-local view is equal to one,* i.e. $\alpha^*(1) = 1$, **if and only if** *all the receivers have degree* 1.

*Proof:* We will first show that in a Z-network network, $\alpha^* \leq 1/2$.

For a deterministic network model, assume that a normalized sum rate of $\alpha$ is achievable; then
$$R_i \geq \alpha n_{ii} - \tau, \ \forall \ 1 \leq i \leq 2. \quad (7)$$

When all the channel gains are $n$, the condition that data can be decoded at the intended destinations gives
$$R_1 + R_2 \leq n.$$

Thus,
$$\alpha(2n) - 2\tau \leq R_1 + R_2 \leq n,$$
or,
$$(2\alpha - 1)n \leq 2\tau.$$

Since this has to hold for all values of $n$ where $\alpha$ and $\tau$ are independent of $n$, $\alpha \leq 1/2$.

For a Gaussian network model, if a normalized sum rate of $\alpha$ is achievable; then
$$R_i \geq \alpha \log(1 + |h_{ii}|^2) - \tau, \ \forall \ 1 \leq i \leq 2. \quad (8)$$

Further, when all $h_{11} = h_{12} = h_{22}$,
$$R_1 + R_2 \leq \log(1 + 2|h_{11}|^2)$$

This gives
$$(2\alpha - 1)\log(1 + |h_{11}|^2) \leq 1 + 2\tau \quad (9)$$

Since this has to hold for all values of $|h_{11}|$ where $\alpha$ and $\tau$ are independent of $h$, $\alpha \leq 1/2$.

This shows that for a Z-cnetwork, $\alpha^*(1) \leq 1/2$. If there is a network containing a link from $\mathsf{T}_i$ to $\mathsf{D}_j$ for $i \neq j$, then as a genie consider a system of two users $i$ and $j$ where all other links are 0 and known to all. In this two user system, Z-network will be an outer bound and thus $\alpha^*(1) \leq 1/2$. This proves that if there is a link from $\mathsf{T}_i$ to $\mathsf{D}_j$ for $i \neq j$, $\alpha^*(1) \leq 1/2$; thus proving the theorem. ∎

Thus, with 1-local view, the only network that can support $\alpha^*(1) = 1$ is the one where no transmitter interferes with other receivers, i.e, a network with $K$ completely isolated flows. As a result, for a two-user interference network where transmitters *can* cause interference at other receivers, MIG scheduling will require the two flows to operate in a TDMA fashion. This is because the transmitters do not know any of the interfering link gains and thus have to optimize for the worst case in our formulation. The worst case network conditions are when the interfering channel gains are the same as the direct link ($h_{12} = h_{11} = h_{22}$), where the network has only one degree of freedom and each node can thus transmit only half the time [8]. Thus, for the two-user case, the above conclusion can be derived from the results in [8]. Theorem 2 is a generalization to arbitrary $K$-user interference network.

We next provide the characterization of independent subgraphs for two-local view, $h = 2$.

**Theorem 3** (2-local View Independent Subgraphs). *The normalized sum-capacity of a $K$-user interference network (deterministic or Gaussian) with 2-local view is equal to one (*i.e. $\alpha^*(2) = 1$*) if and only if all the connected components are of one of the following forms:*

1) *a one-to-many interference network*
2) *a fully-connected interference network*

*Proof:* A fully-connected network implies all nodes are within two-hops from each other. Thus, in this case, the diameter of such a network is two and thus $h = 2$ constitutes global knowledge. By the definition of normalized sum-rate, $\alpha^*(2) = 1$ for a fully-connected subnetwork.

The proof for the case when the connected component is one-to-many interference network is provided in Appendix A. Further, a converse to the statement of the Theorem is also provided in Appendix A. The result was partially presented at [9] for deterministic network and is extended in this paper by providing outer bounds on $\alpha^*$ for all the three-user topologies along with the Gaussian extensions. ∎

Contrasting Theorems 2 and 3, we see that increasing the local horizon from $h = 1$ to $h = 2$ increases the number of networks under which universally optimal performance can be obtained. While for $h = 1$, universal optimality required no simultaneous transmissions, the independent subgraphs for $h = 2$ constitute a richer class. Not only are the fully connected interference networks possible (since their diameter is 1 for $K = 1$ and 2 for $K \geq 2$), one-to-many subgraphs are also possible even though their diameter is 4 for $K \geq 3$. For one-to-many subgraphs, the interfering transmitter is two hops away from all nodes and thus has full network knowledge. As a result, the optimal strategy is to allow $K - 1$ links to operate at their near-maximum link capacity and for the interfering flow to adjust its rate to cause no harmful interference (either the interference is below the noise floor or completely decodable and thus can be cancelled out). This was proved for two-user chain network in [7], and will be extended to a general $K$ in Appendix A.

### C. An Example

Figure 5(a) gives a case of a six-user 4-to-many interference network. With 1-local view, the MIG Scheduling algorithm can be described as follows. Let $A_1 = \{1\}$, $A_2 = \{2\}$, $A_3 = \{3\}$, $A_4 = \{4\}$, and $A_5 = \{5, 6\}$. Note that we have used a shorthand notation in describing these sets; $A_1 = \{a, b\}$ represents that $A_1$ is subgraph containing $\mathsf{T}_j, \mathsf{D}_j$ for all $j \in A_1$ and all edges between the members of $A_1$ are also implied by this shorthand notation. We use a five time-slot strategy. In the $i^{th}$ time-slot, users in $A_i$ transmit. MIG Scheduling achieves $\alpha(1) = 1/5$. We will show that this scheduling is optimal in Theorem 4.

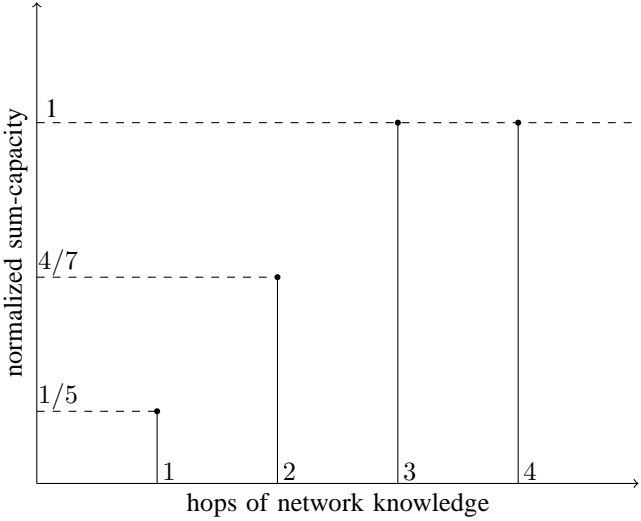

Fig. 6. Normalized sum-capacity vs. $h$-local information for six-user 4-to-many interference network.

With 2-local view, the MIG Scheduling algorithm can be described as follows. Let $A_1 = A_2 = A_3 = \{1,2,3,4\}$, $A_4 = \{1,5,6\}$, $A_5 = \{2,5,6\}$, $A_6 = \{3,5,6\}$, and $A_7 = \{4,5,6\}$. We use a seven time-slot strategy. In the $i^{th}$ time-slot, users in $A_i$ transmit. MIG Scheduling achieves $\alpha(2) = 4/7$. We will show that this scheduling is optimal in Theorem 6. The normalized sum-capacity for increasing local information is depicted in Figure 6.

## IV. Optimality of MIG Scheduling

Now a natural question is: How good is the MIG scheduling? In this section, we address the question and show that MIG scheduling is optimal for several $K$-user networks with 1-local and 2-local view. Our results are limited to 1- and 2-local view only because independent graphs are known only for these two cases.

The reader will immediately note that much like capacity analyses of different multi-terminal networks (multiple access, interference network, etc), our proofs are largely taken on a case by case basis. At the current moment, there appears to be no general algorithmic procedure to derive general capacity region and as a result, we do not have an algorithmic procedure to derive normalized sum-capacity. However, we do note that we can derive normalized sum-capacity in our formulation for many cases while the global-view sum-capacity is still unknown.

### A. 1-local View

Our main result in this section is determining the networks for which MIGS with one-local view is optimal. Recall that one-local view MIGS is equivalent to scheduling of non-interfering links in the network.

The key step in the proof is the derivation of an upper bound. The proof for the upper bound follows the following recipe in all cases for the deterministic model (the Gaussian model is similar).

1) When any transmitter sees the direct channel capacity as $n$, it has to send at a rate $R_i \geq \alpha n - \tau$. This is because if the rate is $< \alpha n - \tau$, then when all other channel gains are 0, the worst-case guarantee of $\alpha$ is not achievable.
2) Find an upper bound on global-view sum capacity when all the channel gains in the network are $n$. Let the global sum capacity be bounded from above by $cn + d$ for some constants $c$ and $d$ which are independent of $n$. For example, one trivial outer bound is $Kn$ for all $K$-user networks. To yield a useful bound, it is important to find the smallest constant $c$.
3) Combining Steps 1 and 2, an outer bound on $\alpha$ as $\alpha \leq c/K$ can be obtained where $K$ is the number of users.

The proof follows the above three steps for each subset of users, and chooses the tightest outer bound thereafter.

Let $A \subseteq \mathsf{G}$ represents a valid interference network with $|A| \leq K$ transmitter-receiver pairs. Suppose the global view sum capacity of $A$ when all the link capacities in $A$ are $n$ is upper bounded by $c_A n + d_A$ for some constants $c_A$ and $d_A$ which may depend on $A$ but remain constant with changing $A$. Then,

$$\alpha \leq \min_A \frac{c_A}{|A|}. \qquad (10)$$

The following theorem characterizes the cases where we can prove that MIG Scheduling is optimal.

**Theorem 4** (1-Local View Optimality of MIG Scheduling). *MIG scheduling is optimal with 1-local view when the network is of one of the following forms, and we also derive $\alpha^*$ for each case.*

1) *All the three user interference networks, except the 3-user cyclic-chain, (In Figure 7, $\alpha^*(1) = 1$ in (a), $\alpha^*(1) = 1/2$ in (b), (c), (d), (e), (f), (g), (j), and (k), and $\alpha^*(1) = 1/3$ in (h), (l), (m), (n), (o), and (p))*
2) *chain interference network, ($\alpha^*(1) = 1/2$ for $K \geq 2$),*
3) *d-to-many interference network, ($\alpha^*(1) = \frac{1}{d+1}$ for $K \geq 2$ and $1 \leq d < K$),*
4) *many-to-d interference network, ($\alpha^*(1) = \frac{1}{d+1}$ for $K \geq 2$ and $1 \leq d < K$),*
5) *fully-connected interference network, ($\alpha^*(1) = \frac{1}{K}$),*

*Further, the achievability holds with $\tau = 0$ for both the deterministic and the Gaussian models.*

*Proof:*
1) For a three user interference network, we will consider all the possible networks as shown in Figure 7 up to relabeling of the users. In networks (b), (c), (d), (e), (f), (g), (j), and (k), the same upperbound as that for the Z-network ($\alpha^* \leq 1/2$) holds since the channel gains except those that forms a Z-network can be made 0 and are known to all as a genie (Since there is only 1-local view, existence of zero capacity links do not help get more information about the network). Further, this can be achieved with MIG scheduling with two time-slots. For (h), (l), (m), (n), (o), and (p), consider the topology equivalent to (h) by setting all other network gains to 0 and make this global information. With this, the outer bound for the case (h) holds for all these cases. In the case (h), suppose all the network gains are the same. Then, $D_1$ decodes the message of $T_1$. Thus, $D_2$ will be able to decode the message of $T_1$ as well as $T_2$ since after decoding message of $T_2$ and subtracting, the equivalent signal is the same as that at $D_1$. Similarly, $D_3$ will be able to decode the message of $T_1$, $T_2$ as well as $T_3$ since after decoding the message of $T_3$ and subtracting, the equivalent signal is same as that at $D_2$. Thus, the normalized sum capacity is upper bounded by that of the Multiple Access Network to $D_3$ thus giving $1/3$ as an upper bound. Further, $1/3$ can be achieved using MIG scheduling, scheduling the three users in three different time-slots.

2) The achievability of $1/2$ follows by using two time-slots, scheduling odd numbered users in the first time-slot and even numbered in the second time-slot, while the outer bound for the Z-network also holds here by the same arguments as in the previous part. Thus, $\alpha^*(1) = 1/2$.

3) As an outer bound, consider $d + 1$ users containing the first $d$ users that are interfering at all receivers and the $d + 1^{th}$ user as one other user. Consider the rest of the direct channel gains as 0 and known to all. In this case, it is easy to see that when all the channel gains are equal, $D_{d+1}$ has to decode all the messages, thus upper bounding the normalized sum capacity by that of the Multiple Access Network to this receiver. For achievability, consider MIG scheduling using $d + 1$ time-slots with $A_1 = \{1\}, \cdots, A_d = \{d\}$ and $A_{d+1} = \{d + 1, \cdots, K\}$. Note that this extends the example of a 4-to-many interference network with 6 users with 1-local view provided in Section III-C.

4) As an outer bound, consider $d + 1$ users containing the first $d$ users that are receiving interference from all transmitters and the $d + 1^{th}$ user as one other user. Assume the rest of the direct channel gains are 0 and known to all. In this case, it is easy to see that when all the channel gains are equal, $D_1$ has to decode all the messages, thus upper bounding the normalized sum capacity by that of the Multiple Access Network to this receiver. For achievability, consider MIG scheduling using $d + 1$ time-slots with $A_1 = \{1\}, \cdots, A_d = \{d\}$ and $A_{d+1} = \{d + 1, \cdots, K\}$.

5) When all the channel gains are equal, each destination has to decode all the messages and is thus upper-bounded by that of the Multiple Access Network, giving $1/K$ as the upper bound. This is achievable using MIG scheduling, scheduling each user in a separate time-slot. ∎

Thus, maximal scheduling of non-interfering links can be information-theoretically optimal for many networks. The theorem only gives sufficient conditions and thus not a sharp characterization of all networks which can be operated optimally with scheduling. However, observing the class of networks given in the theorem, it appears that MIG scheduling might be optimal for a large class of networks. We, thus, explore the connection further in the next section and also discuss the relationship with graph coloring.

*B. 1-Local View: Relation to Maximal Independent Set Scheduling*

For one-local view, the MIG scheduling strategy reduces to maximal independent set scheduling (MIS scheduling) that can be described as follows. An independent set $A_i \subseteq \{1, \cdots, K\}$ is a set that contains mutually non-interfering nodes. A maximal independent set (MIS) is an independent set $A_i$ such that $A_i \cup \{x\}$ is not an independent set for any $x \in \{1, \cdots, K\} \setminus A_i$. Using $t$ time-slots, a maximal independent set $A_i$ is scheduled in each time-slot such that

$$\min_i \frac{1}{t} \sum_{j=1}^{t} \mathbf{1}_{i \in A_j}$$

is maximized over the choice of $t$ and $A_1 \cdots, A_t$. When a user is scheduled, it sends at the direct channel rate (and uses power of 1 in a Gaussian network). The resulting strategy achieves a normalized sub-rate of $\alpha = \min_i \frac{1}{t} \sum_{j=1}^{t} \mathbf{1}_{i \in A_j}$.

This is similar to the following vertex coloring algorithm. To relate to vertex coloring, we will need the concept of conflict graph [4, Chapter 2.2] derived from G as follows. Consider a graph C with $K$ vertices (half as many as present in G), where two vertices $i$ and $j$ are connected if there is an edge between $T_i$ and $D_j$ or between $T_j$ and $D_i$ in G. Suppose that there are $t$ colors, labeled $1, 2, \cdots, t$. We assign $k \leq t$ of these colors to each vertex in C such that the sets of colors associated with two vertices connected by an edge are disjoint. In conventional graph coloring [5], each vertex has only one color and the objective is to assign a color to each vertex such that adjoining vertices have different colors. In contrast, the generalized *set* coloring algorithm can assign multiple colors to each vertex as long as the color sets for adjoining vertices are disjoint. This is similar to fractional coloring considered in [12]. The best set coloring corresponds to MIS schedule and maximizes $k/t$ with $k$ and $t$ as variables. The scheduling algorithm uses $t$ time-slots and schedules the vertices with color $i$ in the $i^{th}$ time-slot.

This algorithm is similar to Maximal Weight Independent Set Scheduling in [3] except that the weights are decided not by the queue lengths, but by the weights that maximize the minimum proportion each link is used.

A $k$-fold coloring of a graph is an assignment of sets of size $k$ to vertices of a graph such that adjacent vertices receive disjoint sets. A $t : k$-coloring is a $k$-fold coloring out of $t$ available colors. The $k$-fold chromatic number $\xi_k$ is the least $t$ such that a $t : k$-coloring exists. Note that MIS Scheduling achieves $\alpha = \max_{k \in \mathbb{N}} \frac{k}{\xi_k}$, where $\xi_k$ is the $k$-fold chromatic number of the conflict graph. The following theorem gives an optimality condition of MIS Scheduling algorithm in terms of the $k$-fold chromatic number of the conflict graph.

**Theorem 5** (1-Local View Optimality of MIS Scheduling). *If the conflict graph of an interference network has $k$-fold*

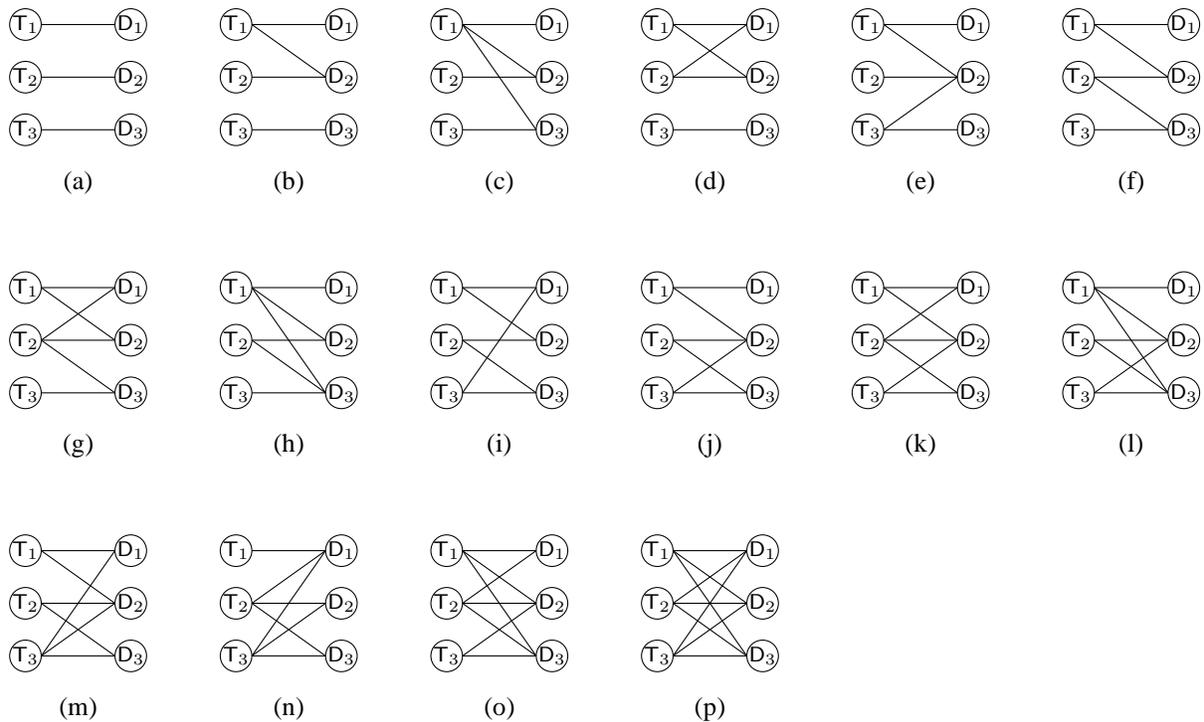

Fig. 7. All possible canonical network topologies in a three-user interference network.

*chromatic number of at most $2k$ for some $k \in \mathbb{N}$, then the MIS scheduling algorithm is optimal, i.e. achieves normalized sum-capacity with 1-local view.*

*Proof:* If the $k$-fold chromatic number in the conflict graph is $k$, then there is no link between any $\mathsf{T}_i$ and $\mathsf{D}_j$ for $j \neq i$. Since this connectivity satisfies the condition of $\alpha^*(1) = 1$, the theorem holds.

If the $k$-fold chromatic number in the conflict graph $k < \xi_k \leq 2k$, then there is at-least one link between $\mathsf{T}_i$ and $\mathsf{D}_j$ for some $j \neq i$. In this case, $\alpha^*(1) \leq 1/2$ by the same arguments as in Theorem 2. Further, this can be achieved by MIS scheduling. ■

**Corollary 1.** *Chain interference networks, different configurations of two-user interference networks, 1-to-many and many-to-1 interference networks are some special cases that have chromatic number $\leq 2$ in the conflict graph. Moreover, the normalized sum capacity is the inverse of the 1-fold chromatic number of the conflict graph in these cases.*

### C. 2-Local View

We start with a theorem which provides sufficient conditions under with MIG scheduling is two-local view optimal.

**Theorem 6** (2-Local View Optimality of MIG Scheduling). *MIG Scheduling achieves normalized sum-capacity with 2-local view when the network is of one of the following forms. We also derive their normalized sum-capacity.*

1) *Two user interference network, ($\alpha^*(2) = 1$),*
2) *Chain interference network, ($\alpha^*(2) = 2/3$ for $K > 2$),*
3) *d-to-many interference network, ($\alpha^*(2) = \frac{d}{2d-1}$ for $1 \leq d < K$ and $K > 2$),*
4) *many-to-one interference network, ($\alpha^*(2) = \frac{K-1}{2K-3}$ for $K > 2$),*
5) *fully-connected interference network, ($\alpha^*(2) = 1$).*

*Proof:*
1) In this case, there are four configurations formed by existence or non-existence of the cross links and in all these configurations, the result follows from Theorem 3.
2) The outer bound of topology (f) in Appendix A holds in this case by assuming all other channel gains to be 0 and known to all. For achievability, the MIG scheduling algorithm can be described as follows. Let $A_j = \{3i+j : i \in \mathbb{Z}, 3i + 1 \leq K\}$ for $j = 1, 2, 3$. According to the MIG scheduling algorithm, three time-slots are used and users in $A_i$ use a strategy that achieves $\alpha(2) = 1$ in the $i^\text{th}$ time-slot. The MIG scheduling strategy achieves $\alpha(2) = 2/3$.
3) Let $d > 1$ because for $d = 1$ the statement holds by Theorem 3. For the outer bound, consider a $d+1$ user d-to-many interference network. Suppose that there exists a scheme achieving normalized sum capacity of $\alpha$. We first prove the result for the deterministic model. For user $d + 1$, since it does not know any other direct channel gain, it has to use $R_{d+1} \geq \alpha n - \tau$ when it sees that all the channel gains within 2 hops have capacity equal to $n$. Suppose that all other direct links have capacity of $n$ while all other cross links have zero capacity. Then, all $R_i \leq (1 - \alpha)n + \tau$ for $i \in [1, d]$ yielding that sum rate $\leq (d - (d-1)\alpha)n + (K - 2)\tau$. This sum rate has to be at-least $\alpha(dn) - \tau$. Since this holds for all $n$, $\alpha \leq \frac{d}{2d-1}$. Similar proof holds in the Gaussian model as follows. For user $d + 1$, since it does not know any other direct channel gain, it has to use $R_{d+1} \geq \alpha \log(1 + \mathsf{SNR}) - \tau$ when it sees that all the channel gains within 2 hops have channel gain equal to $\sqrt{\mathsf{SNR}}$. Suppose that all other direct links have capacity of $\sqrt{\mathsf{SNR}}$ while all other cross links have zero capacity. Then, all $R_i \leq (1-\alpha)\log(1+\mathsf{SNR}) + \tau + 1$ for $i \in [1,d]$ yielding that sum rate $\leq (d-(d-1)\alpha)\log(1+\mathsf{SNR}) + (K-2)\tau + d$. This sum rate has to be at-least $\alpha(d\log(1+\mathsf{SNR})) - \tau$. Since this holds for all SNR, $\alpha \leq \frac{d}{2d-1}$.

For achievability, consider $2d-1$ time-slots in which the first $d-1$ time-slots only users 1 to $d$ transmit. In the remaining $d$ time-slots one user among the first $d$ and all the users $> d$ transmit making it an equivalent one-to-many configuration (or, $A_1 = \cdots = A_{d-1} = \{1,\cdots,d\}$ and $A_{d-1+j} = \{j, d+1, \cdots, K\}$ for $j = 1, \cdots, d$). Thus, this is MIG scheduling with each user scheduled in $d$ time-slots achieving $\alpha^*(2) = \frac{d}{2d-1}$. Note that this extends the example of a 4-to-many interference network with 6 users with 2-local view provided in Section III-C.

4) Suppose that a normalized sum rate of $\alpha$ can be achieved. We first consider a deterministic model. $R_K > \alpha n_{KK} - \tau$ since the $K^{th}$ user has to send at this rate when all other direct channel gains are 0 and are not known to user $K$. Now, suppose all the channel gains are $n$. In this case, $R_i < (1-\alpha)n + \tau$ for $1 \leq i \leq K-1$. Thus, the sum rate achieved is less than $(K-2)(1-\alpha)n + (K-2)\tau + n$. This sum rate has to be at-least $\alpha(K-1)n - \tau$. Since this has to hold for all $n$, $\alpha \leq \frac{K-1}{2K-3}$. For a Gaussian model, $R_K \geq \alpha \log(1 + |h_{KK}|^2) - \tau$ since the $K^{th}$ user has to send at this rate when all other direct channel gains are 0 and are not known to user $K$. Now, suppose all the channel gains are $\sqrt{\mathsf{SNR}}$. In this case, $R_i \leq (1-\alpha)\log(1+\mathsf{SNR}) + \tau + 1$ for $1 \leq i \leq K-1$. Thus, the sum rate achieved is $\leq (K-2)(1-\alpha)\log(1+\mathsf{SNR}) + (K-2)\tau + n + K - 1$. This sum rate has to be at-least $\alpha(K-1)\log(1+\mathsf{SNR}) - \tau$. Since this has to hold for all SNR, $\alpha \leq \frac{K-1}{2K-3}$.

For achievability, consider the data transfer over $2K-3$ time slots. In the timeslot $i$ satisfying $1 \leq i \leq K-1$ users $i$ and $K$ transmit. They form a Z-network and use the optimal strategy for this channel with partial information. In the remaining $K-2$ timeslots, users $1, \cdots, K-1$ transmit at full rate. Let $(R_1, R_2, \cdots, R_K)$ be any point in the global information capacity region. In the $i^{th}$ time-slot where $1 \leq i \leq K-1$, sum rate of at least $R_i + R_K$ can be achieved while in the remaining $K-2$ timeslots, sum rate of at least $\sum_{1 \leq i \leq K-1} R_i$ can be achieved. Thus, the sum-capacity with a factor of $\frac{K-1}{2K-3}$ can be achieved.

5) In this case, the condition of $\alpha^* = 1$ is satisfied by Theorem 3 thus proving the statement.

∎

Note that for all the cases in the statement of Theorem 6, we have characterized normalized sum-capacity in the case of 1-local and 2-local view. For a fully-connected interference network, larger subgraphs increased $\alpha^*(2) = 1$ from $\alpha^*(1) = 1/K$. For a $d$-to-many interference network, one-to-many configurations that satisfy $\alpha^*(2) = 1$ could be exploited to get $\alpha^*(2) = \frac{d}{2d-1}$ from $\alpha^*(1) = 1/2$. With one-local view, only single user encoding and decoding operations are performed while with 2-local view, optimal encoding and decoding operations for one-to-many interference network and fully-connected network need to be performed.

We consider the sixteen network configurations shown in Figure 7 for the two-local view separately in the following theorem. The next theorem shows that MIG Scheduling is normalized sum-capacity achieving for 12 out of 16 canonical cases.

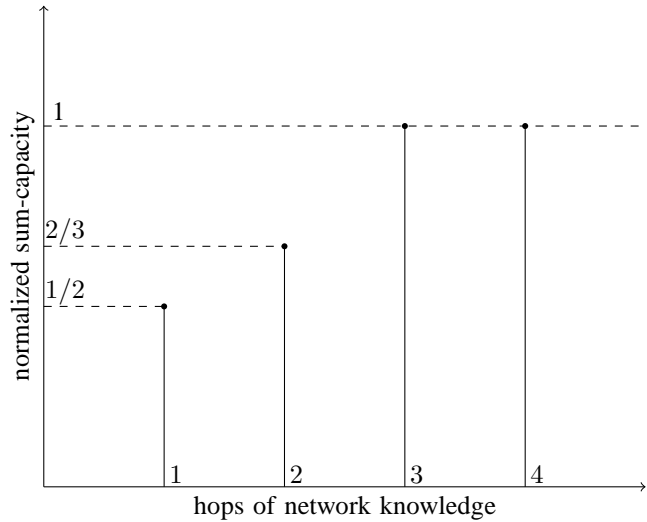

Fig. 8. Normalized sum-capacity vs. $h$-local information for cases (e), (f), (h), (i), (j), (m), (n) in Figure 7.

**Theorem 7** (2-Local View Optimality of MIG Scheduling in Three-user Interference Network). *MIG Scheduling is optimal with 2-local view when the three-user interference network is one of the following types in Figure 7: (a), (b), (c), (d), (e), (f), (h), (i), (j), (m), (n), (p).*

*Proof:* For cases (a), (b), (c), (d), and (p), $\alpha^* = 1$ by Theorem 3. For the remaining cases, the outer bounds of $2/3$ hold as shown in Appendix A. The achievability follows by choosing $A_1 = \{1,2\}$, $A_2 = \{2,3\}$, and $A_3 = \{1,3\}$. The normalized sum-capacity with $h$-local view for varying $h$ in these remaining cases is depicted in Figure 8 ∎

Here, we do not prove the optimality of MIG scheduling for the remaining four cases. We conjecture that the outer bound is tight in cases (g) and (k). The achievability would require the capacity region in these cases to give better schemes, and is left as future work.

## V. Optimality of MIG Scheduling: Extension of MIG Scheduling with 1-local View

Is MIG scheduling always optimal? In this section, we will illustrate an example where MIG scheduling is not optimal. This example will use 1-local view and achieve a normalized sum capacity better than MIS scheduling (MIG scheduling with 1-local information). This gives a way to extend the MIS scheduling with 1-local information to involve coding across the time-slots and hence we define a new strategy called Coded Set (CS) scheduling. This will be followed by some cases when this algorithm is optimal. Finally, we will find normalized sum-capacity for linear deterministic interference networks.

### A. An Example Where MIS Scheduling is not Optimal

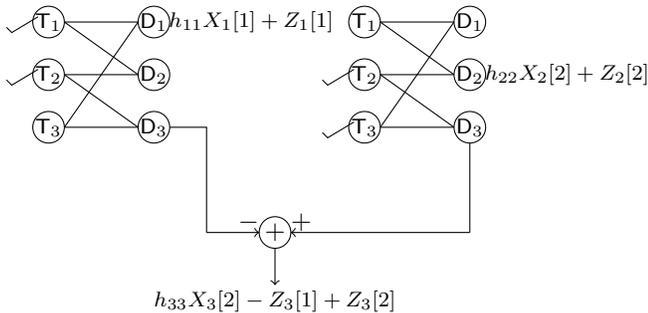

Fig. 9. Two time-slots for CS scheduling. The transmitters with a tick sign transmit, the second user repeats $X_2$ ($X_2[1] = X_2[2]$) in the two time-slots.

We will now illustrate the only case when MIS Scheduling is not optimal in a 3-user interference network, which is a cyclic-chain interference network. The MIS Scheduling algorithm uses three time-slots, scheduling user $i$ in time-slot $i$. Thus, MIG Scheduling achieves $\alpha(1) = 1/3$ (Note that there are only 3 independent sets consisting of individual users and thus optimality of $1/3$ using MIS scheduling is straightforward). We will now describe another strategy for this example, which uses two time-slots as follows (and depicted in Figure 9). The main idea is to perform coding across time. In the first time-slot, we schedule $A_1 = \{1, 2\}$ and in second time-slot, we schedule $A_2 = \{2, 3\}$ such that the codeword of the second user is repeated in the two time-slots. All the users send at the rate equal to the direct link capacity to the intended receiver ($n_{ii}$ in the deterministic and $\log(1 + |h_{ii}|^2)$ in the Gaussian model). In the Gaussian model, power of 1 is used at the first two transmitters while power 2 is used for the third transmitter. Note that this does not effect average power since this transmitter will be used half the time. We will now show that the data can be decoded at the intended receivers. The first receiver can decode its data in the first time-slot since it receives no interference. The second receiver can similarly decode the data in the second time-slot. The third receiver on the other hand subtracts the data received in the first time-slot from that in the second time-slot which gives a interference-free direct signal which can be decoded; double power level at the third transmitter is used since the noise power will also be double the single slot noise power. Thus, all the receivers can decode the data and this strategy achieves $\alpha(1) = 1/2$.

This example motivates an extension of the MIS Scheduling algorithm to involve coding. This new scheduling algorithm is called Coded Set Scheduling (CS Scheduling) which will be described in the next subsection.

### B. Definition of CS Scheduling

In this subsection, we will define the CS algorithm for the deterministic model and the Gaussian model separately. In this section, we will only consider subgraphs $A \subseteq G$ with a set of transmitters $\mathsf{T}_i$ and all the receivers $\{\mathsf{D}_1, \cdots, \mathsf{D}_K\}$ in the subgraph because we do not want to throw away any received signal. Let the in-degree at $\mathsf{D}_i$ be denoted by $d_i$. Suppose that each transmitter generates $k$ independent codewords (The rate of these codewords will be $n_{ii}$ for the deterministic model, and $\log(1 + |h_{ii}|^2/b_i)$ for the Gaussian model where $b_i$ will be defined in the Gaussian subsection below). Let $S_{i,j}$ be a vector of time-slots in which transmitter $\mathsf{T}_i$ is transmitting the $j^{th}$ codeword. Note that each time-slot should be used at a transmitter $\mathsf{T}_i$ for only one codeword, thus giving $S_{i,u}$ and $S_{i,v}$ disjoint for $u \neq v$. Thus, in time-slot $u$, the subgraph $A_u$ used has transmitters $\mathsf{T}_i$ where $i$ satisfies $S_{i,j} \supseteq \{u\}$ for some $1 \leq j \leq k$. The sets $S_{ij}$ and thus $A_u$, $t$ and $k$ are all design variables for the CS scheduling algorithm that satisfy some conditions on the constraint matrix, which is defined next.

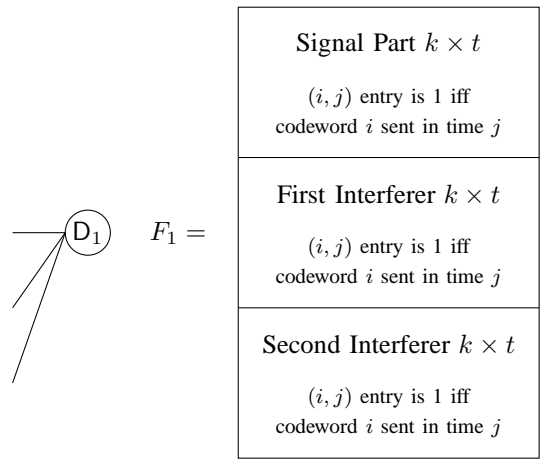

Fig. 10. Constraint matrix with $t$ columns $\mathsf{M}_{1,1}, \cdots \mathsf{M}_{1,t}$ where each column represents the different codewords being sent for the direct signal and the interferers.

We form a binary constraint matrix $F_i$ of size $kd_i \times t$ at each receiver $i$ which is defined as follows. The constraint matrix has $d_i$ blocks of size $k \times t$ where the top block corresponds to the transmitted signal from $\mathsf{T}_i$ while the rest belong to the different transmitters causing interference at $\mathsf{D}_i$ as depicted in Figure 10. In each $k \times t$ subpart of this matrix, only the entries $(j, S_{i,j})$ are 1 for all $1 \leq j \leq k$. Suppose that the $t$ columns of the constraint matrix are denoted as $\mathsf{M}_{i,1}, \cdots \mathsf{M}_{i,t}$ respectively. The constraints that the constraint matrix has to satisfy are different for linear deterministic and Gaussian network model, and is explained below separately for the two cases.

*1) Linear Deterministic Model:* For a linear deterministic network model, CS Scheduling can be described as follows. Suppose that a $kd_i \times t$ matrix with the top $k \times k$ part as an identity and rest of the elements 0 can be formed by choosing each column $j$ as $\sum_{l=1}^{t} a_{jl} \mathsf{M}_{i,l}$ where $a_l$'s are binary and addition is binary addition. If such a transformation exist at destination $i$, this configuration is feasible at vertex $i$. If the assignment of $S_{ij}$ is feasible at each vertex, this strategy achieves $\alpha$ of $k/t$. The strategy that achieves the maximum $k/t$ is called Coded Set (CS) Scheduling.

The scheduling algorithm uses $t$ time-slots. Each user forms $k$ independent codewords at rate $n_{ii}$. User $i$ transmits codeword $j$ in time-slots corresponding to $S_{i,j}$. It is easy to see that the data can be decoded at the receivers. The constraint matrix reduction represents that all the $k$ independent codewords can be decoded in the presence of the interference from other transmitters.

*2) Gaussian Model:* For a Gaussian network model, CS Scheduling can be described as follows. Suppose that a $kd_i \times t$ matrix with the top $k \times k$ part an identity and rest of the elements 0 can be formed by choosing each column $j$ as $\sum_{l=1}^{t} a_{jl} \mathsf{M}_{i,l}$ where $a_l \in \mathbb{R}$ and addition is real addition. If such a transformation exist at destination $i$, this configuration is feasible at vertex $i$. If the assignment of $S_{ij}$ is feasible at each vertex, this strategy achieves $\alpha$ of $k/t$. The strategy that achieves the maximum $k/t$ is called Coded Set (CS) Scheduling.

Note that $a_{jl}$ can be chosen to be 0 for $j > k$. The matrix formed by $a_{jl}$ satisfies

$$\begin{bmatrix} \mathsf{M}_{i,1} & \cdots & \mathsf{M}_{i,t} \end{bmatrix} \begin{bmatrix} a_{11} & \cdots & a_{k1} & 0 & \cdots & 0 \\ \vdots & \ddots & \vdots & \vdots & \ddots & 0 \\ a_{1t} & \cdots & a_{kt} & 0 & \cdots & 0 \end{bmatrix}$$
$$= \begin{bmatrix} I_k & 0 \\ 0 & 0 \end{bmatrix} \quad (11)$$

Since this is an under-determined system, the following is a solution.

$$\begin{bmatrix} a_{11} & \cdots & a_{k1} & 0 & \cdots & 0 \\ \vdots & \ddots & \vdots & \vdots & \ddots & 0 \\ a_{1t} & \cdots & a_{kt} & 0 & \cdots & 0 \end{bmatrix}$$
$$= \begin{bmatrix} \mathsf{M}_{i,1} & \cdots & \mathsf{M}_{i,t} \end{bmatrix}^{\dagger} \begin{bmatrix} I_k & 0 \\ 0 & 0 \end{bmatrix}, \quad (12)$$

where $A^{\dagger}$ represents pseudo-inverse of matrix $A$. Let $b_i = \max_{l=1}^{k} \sum_{m=1}^{t} a_{lm}^2$, where $i$ represents that the constraint matrix is formed for receiver $i$.

User $i$ forms $k$ independent codewords at rate $\log(1 + \mathsf{SNR}_i/b_i)$. User $i$ transmits codeword $j$ in the time-slots in $S_{i,j}$ with power $\mathsf{SNR}_i$. It is easy to see that the data can be decoded at the receivers and the strategy would achieve $\alpha = k/t$ with $\tau \leq \alpha \sum_{i=1}^{K} \log(b_i)$ which is independent of channel gains. This can be further optimized in certain cases by changing the corresponding rates and powers. The constraint matrix specifies the interference and the data at each user; existence of $a_{ij}$'s represents that the data can be decoded in the presence of interference. The value of $b_i$ represents that while decoding the data, the noise gets added up which has to be compensated by the decrease in rate. Since this rate gap is not a function of channel gains, we get a constant $\tau$ that is independent of the channel gains.

*C. Optimality of CS Scheduling*

In this subsection, we prove that CS Scheduling is optimal in a $K$-user cyclic chain, while MIS scheduling is not in an odd user cyclic chain.

**Theorem 8** (1-local View Optimality of CS Scheduling). *CS scheduling is optimal with 1-local view for a $K$-user cyclic chain, while MIS is not for odd $K \geq 3$. Further, an achievable strategy with $\tau = 0$ is possible in this case for both the deterministic and the Gaussian model.*

*Proof:* For a $K$-user cyclic chain, an outer bound of $1/2$ holds from the Z-network by the same arguments as in Theorem 2. If the number of users is even, then using two time-slots and scheduling even and odd users as in MIS scheduling will be optimal. If the number of users is odd, consider two time-slots. In the first time-slot, all odd users transmit while in the even time-slot all even users transmit except that the last user also transmits but it repeats the data in the previous time-slot.

In a deterministic model, all the users send at full rate ($n_{ii}$ for $\mathsf{T}_i$) and the data can be proved to be decoded.

For a Gaussian network, user $i$ sends at a rate of $\log(1 + |h_{ii}|^2)$. All the users except the first use power of 1 to send the data while the first transmitter uses power of 2.

The data can be decoded in the same way as explained for a 3-user cyclic chain thus proving the result. ∎

*D. Normalized Sum-Capacity of Linear Deterministic Networks with 1-local View*

We noted that MIS is not always optimal. Thus, a question arises as to what is the optimal algorithm with 1-local view. In this subsection, we answer this question for linear deterministic interference networks. We first start with some definitions.

**Definition 9.** *A binary model of a given interference network is a linear deterministic model with channel gains of links in $\mathsf{E}$ equal to 1 and all the rest equal to 0.*

**Definition 10.** *Symmetric capacity of an interference network is the maximum $r$ such that rate pair $(r, r, \cdots, r)$ is in the capacity region of the interference network with the full information of network and channel gains.*

We will now show that the normalized sum-capacity of any linear deterministic interference network with 1-local view is the symmetric capacity of the binary model of that interference network in the following theorem.

**Theorem 9** (Normalized Sum-Capacity of Linear Deterministic Networks with 1-local View). *The normalized sum-capacity of any linear deterministic interference network with 1-local view is the symmetric capacity of the binary model of that interference network.*

*Proof:*

Let $C_s$ be the symmetric capacity of the binary model of that interference network, and $\alpha^*$ be the normalized sum-capacity of the interference network.

We will first show that $\alpha^* \leq C_s$. Suppose that a transmitter sees the direct channel gain to its intended receiver is $m$. Then, the rate chosen by the transmitter is at-least $\alpha^* m - \tau$. When all the links in the interference network have gain of $m$, each user transmits at a rate $R_i \geq \alpha^* m - \tau$. Since all the channel gains are $m$, the symmetric rate of $\alpha^* - \tau/m$ should be achievable on the binary model for the interference network (since what is done on $m$ levels can be done on 1 level through time-extension). Since $m$ is arbitrary and thus can be taken to be large enough, the rate pair $(\alpha^*, \alpha^*, \cdots, \alpha^*)$ should be in the capacity region for the binary model for the interference network which gives $\alpha^* \leq C_s$.

We will now show that $\alpha^* \geq C_s$ which will prove the statement of the theorem. To prove this, we will use the optimal strategy for the binary model that achieves symmetric capacity $C_s$ and use it in the original linear deterministic interference network. We use the symmetric capacity achieving scheme for the binary model of interference network (that achieves symmetric capacity $C_s$) at all the bit levels of the original linear deterministic interference network. The sum rate achieved is at-least $\alpha$ times the sum of all direct channel capacities and hence normalized sum rate of $\alpha$ is achievable. We only need to prove that the data can be decoded at the receivers. To see this, note that every receiver is receiving at-most the same interference as in the case of the binary model and hence can fake other interference and still decode the data. ■

## VI. BEYOND 2-LOCAL VIEW

The problem of finding independent sub-graphs is open in general, while we have provided cases for $h = 1$ or $2$. In this section, we will see some cases when $\alpha^*(3) = 1$.

**Theorem 10** (3-local View Independent Subgraphs). *Normalized sum-capacity of a $K$-user interference network with 3-local view is equal to one (i.e. $\alpha^*(3) = 1$) if the network has all its connected components satisfying any of the following*

1) *many-to-$d$ interference network*
2) *$d$-to-many interference network*
3) *All configurations in three user interference networks in Figure 7 except for (g).*

*Proof:* In all the cases except cases (f) and (g) in Figure 7, 3-local view is the global view. Thus, we only need to show the result in case (f). The capacity region in this case is known exactly for deterministic region [7] which will be used to prove this case.

The deterministic network capacity region for a three user double Z interference network is the set of nonnegative rates $(R_1, R_2, R_3)$ satisfying [7]

$$\begin{aligned} R_i &\leq n_{ii}, \, i = 1, 2, 3 \\ R_1 + R_2 &\leq \max(n_{11}, n_{12}, n_{22}, n_{11} + n_{22} - n_{12}) \\ R_2 + R_3 &\leq \max(n_{22}, n_{23}, n_{33}, n_{22} + n_{33} - n_{23}) \\ R_1 + R_2 + R_3 &\leq \max(n_{33}, n_{23}) + (n_{11} - n_{12})^+ \\ &\quad + \max(n_{12}, n_{22} - n_{23}). \end{aligned}$$

Since all the transmitter have 3-hops of channel gain information, we will use the following strategy. The first and the third transmitter send at rate $n_{ii}$ while the second transmitter knows all the channel gains and thus backs off so that receiver 2 can decode the data and receiver 3 is also able to decode.

1) $n_{12} \leq n_{11}$: In this case, the second user do not transmit on the lower $n_{12}$ levels and uses a strategy for the lower Z-network consisting of the second and the third user with equivalent channel gain between the second transmitter-receiver pair being $n_{22} - n_{12}$. Thus, the following sum rate can be achieved.

$$\begin{aligned} R_{ach} &= n_{11} + \max\left(n_{22} - n_{12}, n_{33}, \right. \\ &\quad \min(n_{23}, n_{22} + n_{33} - n_{12}), \\ &\quad \left. n_{22} + n_{33} - n_{23} - n_{12}\right). \end{aligned}$$

Using the above capacity region, this achievable sum-rate can also be shown to be optimal.

2) $n_{12} \geq n_{11}$ and $n_{22} \leq n_{12} - n_{11}$: In this case, the second transmitter does not receive any interference and thus the sum rate of $n_{11} + \max(n_{22}, n_{33}, \min(n_{23}, n_{22} + n_{33}), n_{22} + n_{33} - n_{23})$ can be achieved which is also optimal.

3) $n_{12} \geq n_{11}$, $n_{12} - n_{11} \leq n_{22} \leq n_{12}$, and $n_{23} \leq n_{33}$: In this case, the second transmitter sends at lower $\min(n_{12} - n_{11}, (n_{22} - n_{23})^+)$. This is also optimal.

4) $n_{12} \geq n_{11}$, $n_{12} - n_{11} \leq n_{22} \leq n_{12}$, and $n_{23} > n_{33}$: In this case, the second user can send data on the levels which are interfering at the second receiver such that they are also repeated in the lower $n_{12} - n_{11}$ levels. Thus, rate of $\min((n_{12} - n_{11}), (n_{22} - n_{23})^+ + \min(n_{22}, n_{23} - n_{33}))$ can be supported at the second transmitter.

5) $n_{12} \geq n_{11}$, $n_{22} \geq n_{12}$, $n_{23} \leq n_{33}$, and $n_{23} \leq n_{22} - n_{12}$: In this case, the second transmitter does not send at interfered $n_{11} + n_{23}$ levels and sends at rate $(n_{22} - n_{11} - n_{23})^+$.

6) $n_{12} \geq n_{11}$, $n_{22} \geq n_{12}$, $n_{23} \leq n_{33}$, and $n_{22} - n_{12} \leq n_{23} \leq n_{11} + n_{22} - n_{12}$: In this case, the second transmitter only sends at lower $(n_{12} - n_{11})^+$ levels.

7) $n_{12} \geq n_{11}$, $n_{22} \geq n_{12}$, $n_{23} \leq n_{33}$, and $n_{23} \geq n_{11} + n_{22} - n_{12}$: In this case, the second transmitter sends does not send at $n_{23}$ levels producing interference and hence sends at a rate of $(n_{22} - n_{23})^+$.

8) $n_{12} \geq n_{11}$, $n_{22} \geq n_{12}$, $n_{23} \geq n_{33}$, and $n_{23} \leq n_{22} - n_{12}$: In this case, the second transmitter does not transmit on $n_{11}$ levels at which it receives interference and $n_{33}$ levels at which it causes interference and this use a rate of $n_{22} - n_{11} - n_{33}$.

9) $n_{12} \geq n_{11}$, $n_{22} \geq n_{12}$, $n_{23} \geq n_{33}$, and $n_{22} - n_{12} \leq n_{23} \leq n_{11} + n_{22} - n_{12}$: In this case, the second transmitter sends at lower $n_{12} - n_{11}$ levels and top $\min(n_{23} - n_{33}, n_{22} - n_{12})$ levels.

10) $n_{12} \geq n_{11}$, $n_{22} \geq n_{12}$, $n_{23} \geq n_{33}$, and $n_{22} + n_{11} - n_{12} \leq n_{23} \leq n_{22}$: The second transmitter sends at a rate of $n_{22} - \max(n_{11}, n_{33})$ since some of the levels at which interference is caused at the third receiver can be repeated at the $n_{11}$ levels at which it is receiving interference such that they can be decoded at the third receiver.

11) $n_{12} \geq n_{11}$, $n_{22} \geq n_{12}$, $n_{23} \geq n_{33}$, and $n_{23} \geq n_{22}$: The second transmitter sends at a rate of $n_{22} - \max(n_{11}, (n_{33} + n_{22} - n_{23})^+)$ since $(n_{33} + n_{22} - n_{23})^+$ are the effective levels that the second user produces interference to the third user and the same arguments as in the previous case apply. It can also be shown that this results in the maximum sum rate.

The results can be extended to Gaussian network as in shown in Appendix B. ∎

We see that for a general proof of achievability, we need to find the general capacity region. This is hard in general and is the reason that understanding for general number of hops is open. Note that a graph with very few links and large number of links will be known completely within a small number of hops. However, there are cases in the middle where the capacity region may be required so as to say if $\alpha^* = 1$ or not. For example, in 3-user interference network, we only needed to consider 2 cases for 3-local view. We resolved the case (f), while case (g) is still open. We conjecture that $\alpha^*(3) = 1$ for any 3-user interference network. However to prove this, we need a strategy for (g) which achieves the sum-capacity and is still open.

## VII. CONCLUSIONS

In this paper, we give a framework for optimality of distributed decisions. The optimality is measured in terms of normalized sum-capacity which is the best worst case guarantee of the distributed decisions. We gave an achievable algorithm called maximal independent graph scheduling, and characterized its performance in several examples. We found this algorithm to achieve normalized sum-capacity in several cases, while we also show that this algorithm is not always optimal. We also find the normalized sum capacity of linear deterministic interference networks with 1-local view.

## APPENDIX A
## UNIVERSALLY OPTIMAL STRATEGIES WITH 2 HOPS IN THREE USER TOPOLOGIES

We first note that for $K < 3$, all the topologies have connected components that satisfy the property in the statement of the theorem and thus the result holds trivially.

In a three-user interference network, there are at-most six cross links, existence or non-existence of which gives rise to $2^6 = 64$ cases. Some of the cases are topologically equivalent (up to relabeling of users) and hence that will reduce the total number of possibilities considered in this chapter to the sixteen that are shown in Figure 7.

**(e)**: Suppose that the normalized sum rate of $\alpha$ can be achieved. Further, suppose that the second user sees the channel gains as $n_{22} = n_{12} = n_{32} = n$. In this case, the rate allocated by the second user has to be at-least $\alpha n - \tau$ for some $\tau$ independent of $n$. This is because the achievable sum rate has to be at-least $\alpha C_{sum} - \tau$ even if $n_{11} = n_{33} = 0$. Now suppose all the channel gains are equal to $n$. In this case since $R_1 + R_2 \leq n$, we have $R_1 \leq (1-\alpha)n + \tau$. Further since $R_2 + R_3 \leq n$, we have $R_1 + R_2 + R_3 \leq (2-\alpha)n + \tau$. This sum rate has to be at-least $\alpha(2n) - \tau$ since the sum capacity is $2n$. Thus, $2\alpha n - \tau \leq (2-\alpha)n + \tau$ or $(3\alpha - 2)n \leq 2\tau$. If $3\alpha - 2 > 0$, $n$ can be chosen large enough to not satisfy the inequality. So, the inequality can be satisfied for all $n$ only when $3\alpha - 2 \leq 0$ which gives $\alpha \leq 2/3$.

**(f)**: Suppose that the normalized sum rate of $\alpha$ can be achieved. Further, suppose that the first user sees the channel gains as $n_{11} = n_{12} = n_{22} = n$. In this case, the second user will send at rate $\geq \alpha n - \tau$ if $n_{33} = 0$ by the same arguments as in part (e) which implies $R_1 \leq (1-\alpha)n + \tau$. Now suppose all the channel gains are equal to $n$. In this case since $R_2 + R_3 \leq n$, we have $R_1 + R_2 + R_3 \leq (2-\alpha)n + \tau$. This sum rate has to be at-least $\alpha(2n) - \tau$ since the sum capacity is $2n$. Thus, $2\alpha n - \tau \leq (2-\alpha)n + \tau$ or $(3\alpha - 2)n \leq 2\tau$. Similar arguments as in (e) yields $\alpha \leq 2/3$.

**(g)**: Suppose that the normalized sum rate of $\alpha$ can be achieved. Further, suppose that the first user sees the channel gains as $n_{11} = n_{12} = n_{21} = n$, $n_{22} = 2n$. In this case, $2R_1 + R_2 \leq 2n$ gives that if $R_1 = x$, $R_1 + R_2 \leq 2n - x$. If $n_{33} = 0$, the first user should give a strategy such that $(R_1, R_2)$ satisfy $R_1 + R_2 \geq 2n\alpha - \tau$ giving $R_1 \leq 2n(1-\alpha) + \tau$. Now suppose that $n_{23} = n_{33} = 2n$ giving $R_2 + R_3 \leq 2n$. We thus have $R_1 + R_2 + R_3 \leq 2n(2-\alpha) + \tau$. Since this has to be at-least $3n\alpha - \tau$, using similar arguments as in (e) yields $\alpha \leq 4/5$.

**(h)**: Suppose that the normalized sum rate of $\alpha$ can be achieved. Further, suppose that the third user sees the channel gains as $n_{33} = n_{13} = n_{23} = n$. In this case, the third user will send at rate $\geq \alpha n - \tau$ if $n_{11} = n_{22} = 0$ by the same arguments as in part (e). Further suppose $n_{11} = n_{22} = n$, $n_{12} = 0$ which implies $R_1 \leq (1-\alpha)n + \tau$. Since $R_2 + R_3 \leq n$, we have $R_1 + R_2 + R_3 \leq (2-\alpha)n + \tau$. This sum rate has to be at-least $\alpha(2n) - \tau$ since the sum capacity is $2n$. Thus, $2\alpha n - \tau \leq (2-\alpha)n + \tau$ or $(3\alpha - 2)n \leq 2\tau$. Similar arguments as in (e) yields $\alpha \leq 2/3$.

**(i)**: Suppose that the normalized sum rate of $\alpha$ can be achieved. Further, suppose that the first user sees all the channel gains in two hops equal to $n$. In this case, if $n_{33} = 0$, the second user will have to send at rate $\geq \alpha n - \tau$ and thus $R_1 \leq (1-\alpha)n + \tau$. Further suppose all the channel gains are equal to $n$ which implies $R_2 + R_3 \leq n$ thus giving $R_1 + R_2 + R_3 \leq (2-\alpha)n + \tau$. This sum rate has to be at-least $\alpha(2n) - \tau$ since the sum capacity is $2n$. Thus, $2\alpha n - \tau \leq (2-\alpha)n + \tau$ or $(3\alpha - 2)n \leq 2\tau$. Similar arguments as in (e) yields $\alpha \leq 2/3$.

**(j)**: The same steps as in (i) yield $\alpha \leq 2/3$.

**(k)**: Let $n_{32} = 0$ be the global information. Applying the same steps as in (g) for the other channel gains gives $\alpha \leq 4/5$.

**(l)**: Suppose that the normalized sum rate of $\alpha$ can be achieved. Further, suppose that the second user sees the channel gains as $n_{22} = n_{23} = n_{32} = n$, $n_{33} = 2n$. In this case, we get $R_2 \leq 2n(1-\alpha) + \tau$ as in (g). Now suppose that $n_{13} = n_{33} = 2n$, $n_{12} = 0$ giving $R_1 + R_3 \leq 2n$. Using similar arguments as in (g) yields $\alpha \leq 4/5$.

**(m)**: Let $n_{31} = 0$ be the global information. Applying same steps for the remaining channel gains as in (i) yields $\alpha \leq 2/3$.

**(n)**: Suppose that the normalized sum rate of $\alpha$ can be achieved. Then, if the first user sees all the channel gains as $n$ in two hops, $R_1 \geq \alpha n - \tau$. Suppose that $n_{22} = n_{33} = n$, $n_{23} = n_{32} = 0$. In this case, $R_1 + R_2 + R_3 \leq (2-\alpha)n + \tau$ and since it has to be at-least $2\alpha n - \tau$ for all $n$, $\alpha \leq 2/3$.

**(o)**: Suppose that the normalized sum rate of $\alpha$ can be achieved. Further, suppose that the third user sees the channel gains as $n_{33} = n_{23} = n_{32} = n$, $n_{22} = 2n$, $n_{13} = 0$. In this case, we get $R_3 \leq 2n(1-\alpha) + \tau$ as in (g). Now suppose that $n_{12} = n_{11} = n_{21} = 2n$ giving $R_1 + R_2 \leq 2n$. Using similar arguments as in (g) yields $\alpha \leq 4/5$.

We will now consider $K > 3$. Consider that there exist a connected component with $K > 3$ users which is not in the one-to-many configuration or in the fully-connected configuration. Then, two cases arise:
1) There exists a transmitter (say $T_1$) which has degree $d$ satisfying $1 < d < K$.
2) All the transmitter nodes have degrees 1 or $K$, such that the number of nodes $n$ having degree $K$ satisfy $1 < n < K$.

For the first case, take the nodes $1, \cdots, d$ as the nodes whose receivers are connected to $T_1$. Now, there exist a transmitter-receiver pair among $d+1, \cdots, K$ whose transmitter or receiver is connected to any of the nodes $1, \cdots, d$. Choose any such pair and call it pair $d+1$. The receiver of $d+1$ is not connected to transmitter 1. Now if the receiver of the first node is connected to the transmitter of $d+1$, then choose the nodes $1, 2, d+1$ and assume that the direct link of all other users is zero and this information is given as a genie to all the nodes. This creates a genie-aided system in which the nodes 1, 2 and $d+1$ have the uncertainties about all the links connecting them and know 2-local view among these links only. In this genie-aided system, there does not exist any universally optimal strategy, thus proving the claim (since it makes a connected three-user component which is not in the one-to-many configuration or in the fully-connected configuration). If pair $d+1$ is not connected to pair 1, let us say it is connected to pair $2 \leq j \leq d$. Then, choosing nodes $1, j, d+1$ and repeating the same argument as above proves the statement.

For the second case, choose the three nodes as any two nodes in which the transmitter has degree $K$ and one in which the transmitter has degree 1. Repeating the above genie-aided proof for these three nodes proves the theorem.

This completes the proof that there does not exist a universally optimal strategy for a topology that does contain a connected component which is not in the one-to-many configuration or in the fully-connected configuration.

It is easy to see that there exists a strategy with $\alpha = 1$ if all the connected components of the topology are in one-to-many configuration or fully-connected configuration. For fully-connected components, all the nodes know their connected components and thus each node in the component can use the optimal strategy for its component. For the one-to-many components, each of the users whose transmitters have degree 1 send at rate equal to the rate that the direct channel can support and the remaining user knows all the channel gain and adjust its rate correspondingly. Assume that it is a one-to-many component of $L$ users with the first transmitter having degree $L$. The above strategy achieves a sum rate of $R_{sum} = \sum_{i=2}^{L} n_{ii} + \sum_{i=1}^{n_{11}} \mathbf{1}_{|U_k|=0}$, where $|U_k|$ is the number of users potentially experiencing interference from the $k^{th}$ signal level of first transmitter which is the same as the sum capacity with global channel information in [11].

We now see the steps extended to a Gaussian network model. For this, we only consider case (e). Extension of the remaining steps is similar and is thus omitted.

**(e)**: Suppose that a normalized sum rate of $\alpha$ can be achieved. Further, suppose that the second user sees the channel gains as $h_{22} = h_{12} = h_{32} = \sqrt{\mathsf{SNR}}$. In this case, the rate allocated by the second user has to be at-least $\alpha \log(1+\mathsf{SNR}) - \tau$ for some $\tau$ independent of $n$. This is because the achievable sum rate has to be at-least $\alpha C_{sum} - \tau$ even if $h_{11} = h_{33} = 0$. Now suppose all the channel gains are equal to $\sqrt{\mathsf{SNR}}$. In this case since $R_1 + R_2 \leq \log(1+\mathsf{SNR}) + 1$, we have $R_1 \leq (1-\alpha)\log(1+\mathsf{SNR}) + \tau + 1$. Further since $R_2 + R_3 \leq \log(1+\mathsf{SNR}) + 1$, we have $R_1 + R_2 + R_3 \leq (2-\alpha)n + \tau + 2$. This sum rate has to be at-least $\alpha(2\log(1+\mathsf{SNR})) - \tau$ since the sum capacity is at-least $2\log(1+\mathsf{SNR})$. Thus, $2\alpha \log(1+\mathsf{SNR}) - \tau \leq (2-\alpha)\log(1+\mathsf{SNR}) + \tau + 2$ or $(3\alpha - 2)\log(1+\mathsf{SNR}) \leq 2\tau + 2$. If $3\alpha - 2 > 0$, SNR can be chosen large enough to not satisfy the inequality. So, the inequality can be satisfied for all SNR only when $3\alpha - 2 \leq 0$ which gives $\alpha \leq 2/3$.

For the achievability, sum capacity can be achieved for a fully-connected interference network since every user knows the global network state. For one-to-many network, the result in [11] gives that all users except the first using rate $(\log(\mathsf{SNR}_i))^+$ and the first user backing off will achieve a sum rate within $3K - 2$ bits of the sum capacity.

## APPENDIX B
## PROOF FOR GAUSSIAN NETWORK FOR 3 HOP OPTIMALITY OF Z-CHAIN IN THEOREM

The capacity region for the three-user Z-chain interference network is upper bounded by the following regions [7]. (We will use $|h_{ii}|^2 = \mathsf{SNR}_i$ and $|h_{ij}|^2 = \mathsf{INR}_j$ for $j \neq i$ in this Appendix. Further note that although in [7], it is mentioned that $h_{ii}$ and $h_{ij}$ are positive reals, the outer bound proof extends to general channel gains using same arguments.)

1) $\mathsf{INR}_2 \geq \mathsf{SNR}_1$ and $\mathsf{INR}_3 \geq \mathsf{SNR}_2$: In this case, an outer bound on the rate region is given as follows.

$$R_1 \leq \log(1+\mathsf{SNR}_1) \qquad (14a)$$
$$R_2 \leq \log(1+\mathsf{SNR}_2) \qquad (14b)$$
$$R_3 \leq \log(1+\mathsf{SNR}_3) \qquad (14c)$$
$$R_1 + R_2 \leq \log(1+\mathsf{SNR}_2 + \mathsf{INR}_2) \qquad (14d)$$
$$R_2 + R_3 \leq \log(1+\mathsf{SNR}_3 + \mathsf{INR}_3) \qquad (14e)$$

2) $\mathsf{INR}_2 \geq \mathsf{SNR}_1$ and $\mathsf{INR}_3 \leq \mathsf{SNR}_2$: In this case, an outer bound on the rate region is given as follows.

$$R_1 \leq \log(1+\mathsf{SNR}_1) \qquad (15a)$$
$$R_2 \leq \log(1+\mathsf{SNR}_2) \qquad (15b)$$
$$R_3 \leq \log(1+\mathsf{SNR}_3) \qquad (15c)$$
$$R_1 + R_2 \leq \log(1+\mathsf{SNR}_2 + \mathsf{INR}_2) \qquad (15d)$$
$$R_2 + R_3 \leq \log(1+\mathsf{SNR}_2)$$
$$\qquad + \log(1+\frac{\mathsf{SNR}_3}{1+\mathsf{INR}_3}) \qquad (15e)$$

Further, if $(\mathsf{INR}_2 + 1)\mathsf{INR}_3 \leq \mathsf{SNR}_2$

$$R_1 + R_2 + R_3 \leq \log(1+\frac{\mathsf{SNR}_3}{1+\mathsf{INR}_3})$$
$$\qquad + \log(1+\mathsf{INR}_2 + \mathsf{SNR}_2) \quad (16)$$

else if $(\mathsf{INR}_2+1)\mathsf{INR}_3 \geq \mathsf{SNR}_2$

$$R_1 + R_2 + R_3 \leq \log(1+\mathsf{INR}_3+\mathsf{SNR}_3) \\ + \log(1+\mathsf{INR}_2). \quad (17)$$

3) $\mathsf{INR}_2 \leq \mathsf{SNR}_1$ and $\mathsf{INR}_3 \geq \mathsf{SNR}_2$: In this case, an outer bound on the rate region is given as follows.

$$R_1 \leq \log(1+\mathsf{SNR}_1) \quad (18a)$$
$$R_2 \leq \log(1+\mathsf{SNR}_2) \quad (18b)$$
$$R_3 \leq \log(1+\mathsf{SNR}_3) \quad (18c)$$
$$R_1 + R_2 \leq \log(1+\mathsf{SNR}_1) \\ + \log\left(1+\frac{\mathsf{SNR}_2}{1+\mathsf{INR}_2}\right) \quad (18d)$$
$$R_2 + R_3 \leq \log(1+\mathsf{SNR}_3+\mathsf{INR}_3) \quad (18e)$$

4) $\mathsf{INR}_2 \leq \mathsf{SNR}_1$ and $\mathsf{INR}_3 \leq \mathsf{SNR}_2$: In this case, an outer bound on the rate region is given as follows.

$$R_1 \leq \log(1+\mathsf{SNR}_1) \quad (19a)$$
$$R_2 \leq \log(1+\mathsf{SNR}_2) \quad (19b)$$
$$R_3 \leq \log(1+\mathsf{SNR}_3) \quad (19c)$$
$$R_1 + R_2 \leq \log(1+\mathsf{SNR}_1) \\ + \log\left(1+\frac{\mathsf{SNR}_2}{1+\mathsf{INR}_2}\right) \quad (19d)$$
$$R_2 + R_3 \leq \log(1+\mathsf{SNR}_2) \\ + \log\left(1+\frac{\mathsf{SNR}_3}{1+\mathsf{INR}_3}\right) \quad (19e)$$

Further, if $(\mathsf{INR}_2+1)\mathsf{INR}_3 \leq \mathsf{SNR}_2$

$$R_1 + R_2 + R_3 \leq \log(1+\mathsf{SNR}_1) \\ + \log\left(1+\frac{\mathsf{SNR}_2}{1+\mathsf{INR}_2}\right) \\ + \log\left(1+\frac{\mathsf{SNR}_3}{1+\mathsf{INR}_3}\right) \quad (20)$$

else if $(\mathsf{INR}_2+1)\mathsf{INR}_3 \geq \mathsf{SNR}_2$

$$R_1 + R_2 + R_3 \leq \log(1+\mathsf{SNR}_1) \\ + \log(1+\mathsf{INR}_3+\mathsf{SNR}_3). (21)$$

With this, we will now show that the achievability is within at-most 4 bits from the outer bound. We will assume $R_1 = \log(1+\mathsf{SNR}_1)$, $R_3 = \log(1+\mathsf{SNR}_3\frac{1+\mathsf{INR}_3}{1+2\mathsf{INR}_3})$. Second transmitter will choose a rate backing off to other users as will be shown in the following cases.

*A.* $\mathsf{INR}_2 \leq \mathsf{SNR}_1$

In this case, the second assumes $\mathsf{SNR}'_2 = \frac{\mathsf{SNR}_2}{1+\mathsf{INR}_2}$ and use the strategy for Z-network in [7] consisting of only second and third users. Using this, it is easy to show that the achievable rate is as follows in the two cases as described below.
1) If $\mathsf{INR}_3 \leq \frac{\mathsf{SNR}_2}{1+\mathsf{INR}_2}$, a rate of $\log(1+\mathsf{SNR}_1) + \log(1+\frac{\mathsf{SNR}_2}{1+\mathsf{INR}_2}) + \log(1+\frac{\mathsf{SNR}_3}{1+\mathsf{INR}_3}) - 2$ can be achieved.
2) If $\mathsf{INR}_3 \geq \frac{\mathsf{SNR}_2}{1+\mathsf{INR}_2}$, a rate of $\log(1+\mathsf{SNR}_1) + \log(1+\mathsf{SNR}_3\mathsf{INR}_3) - 2$ can be achieved.

Thus, sum-capacity within 2-bits can be achieved.

*B.* $\mathsf{INR}_2 \geq \mathsf{SNR}_1$, $\mathsf{SNR}_1 \leq \frac{\mathsf{INR}_2}{1+\mathsf{SNR}_2}$

In this case, the second user uses the same strategy as the lower Z-network consisting of second and third users since the destination will be able to decode receiver 1's data treating the data of second user as noise. Thus, sum-capacity within 2 bits can be achieved.

*C.* $\mathsf{INR}_2 \geq \mathsf{SNR}_1$, $\mathsf{SNR}_1 \geq \frac{\mathsf{INR}_2}{1+\mathsf{SNR}_2}$, $\mathsf{SNR}_2 \leq \mathsf{INR}_2$, $\mathsf{INR}_3 \leq \mathsf{SNR}_3$

In this case, the second user makes a codebook of rate $\min(\log(1+\mathsf{INR}_2) - \log(1+\mathsf{SNR}_1), \log(1+\frac{\mathsf{SNR}_2}{1+\mathsf{INR}_3}))$ and transmits it using a power level of $\frac{1}{1+\mathsf{INR}_3}$. This can be decoded jointly with the data from the first transmitter since

$$R_1 \leq \log(1+\mathsf{INR}_2) \quad (22a)$$
$$R_2 \leq \log(1+\frac{\mathsf{SNR}_2}{1+\mathsf{INR}_3}) \quad (22b)$$
$$R_1 + R_2 \leq \log(1+\frac{\mathsf{SNR}_2}{1+\mathsf{INR}_3}+\mathsf{INR}_2) \quad (22c)$$

For the outer bound, if $\mathsf{INR}_3 \geq \mathsf{SNR}_2$, $R_1 + R_2$ is within 1 bit and so is $R_3$ thus giving sum-capacity within 2 bits. If $\mathsf{INR}_3 \leq \mathsf{SNR}_2$, it can be shown that achievability is again within 2 bits in all the cases.

*D.* $\mathsf{INR}_2 \geq \mathsf{SNR}_1$, $\mathsf{SNR}_1 \geq \frac{\mathsf{INR}_2}{1+\mathsf{SNR}_2}$, $\mathsf{SNR}_2 \leq \mathsf{INR}_2$, $\mathsf{INR}_3 \geq \mathsf{SNR}_3$

We divide this case into three sub-cases.
1) $\mathsf{INR}_3 \geq \mathsf{SNR}_2$: In this case, the second transmitter uses a single codebook of rate $\min(\log(1+\mathsf{SNR}_2+\mathsf{INR}_2) - \log(1+\mathsf{SNR}_1), \log(1+\frac{\mathsf{INR}_3}{1+\mathsf{SNR}_3}))$ and sends at power of 1. In this case, the data can be jointly decoded at the second and the third destination. Note that sum capacity can be achieved within 1 bit in this case.
2) $\mathsf{INR}_3 \leq \mathsf{SNR}_2$, $\mathsf{SNR}_1(1+\frac{\mathsf{SNR}_2}{1+\mathsf{INR}_3}) \leq \mathsf{INR}_2$: In this case, the second transmitter forms a single code-book of rate $\log(1+\frac{\mathsf{SNR}_2}{1+\mathsf{INR}_3})$ and sends at a power of $\frac{1}{1+\mathsf{INR}_3}$. The second receiver does joint decoding while the third receiver treats the message from the second transmitter as noise. This achieves the sum capacity within 1 bit.
3) $\mathsf{INR}_3 \leq \mathsf{SNR}_2$, $\mathsf{SNR}_1(1+\frac{\mathsf{SNR}_2}{1+\mathsf{INR}_3}) \geq \mathsf{INR}_2$: In this case, second transmitter forms a single code-book of rate $(\log(\frac{\mathsf{INR}_2}{\mathsf{SNR}_1}))^+$ and sends at a power of $\frac{1}{\mathsf{SNR}_2}(\frac{\mathsf{INR}_2}{\mathsf{SNR}_1}-1)$. The second receiver does joint decoding while the third receiver treats the message from the second transmitter as noise. This achieves the sum-capacity within 2 bits.

*E.* $\mathsf{INR}_2 \geq \mathsf{SNR}_1$, $\mathsf{SNR}_1 \geq \frac{\mathsf{INR}_2}{1+\mathsf{SNR}_2}$, $\mathsf{SNR}_2 \geq \mathsf{INR}_2$, $\mathsf{INR}_3 \leq \mathsf{SNR}_3$, $(\mathsf{INR}_2+1)\mathsf{INR}_3 \leq \mathsf{SNR}_2$

In this case, the second transmitter forms a single code-book of rate $\log(1+\mathsf{INR}_2+\mathsf{SNR}_2) - \log(1+\mathsf{INR}_3) - \log(1+\mathsf{SNR}_1)$ and sends at a power of $\frac{1}{1+\mathsf{INR}_3}$. The second receiver does joint decoding while the third receiver treats the message from the second transmitter as noise. This achieves the sum capacity within 2 bits.

*F.* $\text{INR}_2 \geq \text{SNR}_1$, $\text{SNR}_1 \geq \frac{\text{INR}_2}{1+\text{SNR}_2}$, $\text{SNR}_2 \geq \text{INR}_2$, $\text{INR}_3 \leq \text{SNR}_3$, $(\text{INR}_2+1)\text{INR}_3 \leq \text{SNR}_2$, $\text{SNR}_1(1+\frac{\text{SNR}_2}{1+\text{INR}_3}) \geq \text{INR}_2$

If $\text{INR}_3 \geq \text{SNR}_2$, the second transmitter turns off and achieves within 2 bits of the sum capacity. So, we will only consider the case when $\text{INR}_3 \leq \text{SNR}_2$. In this case, the second transmitter forms a single code-book of rate $\log(\frac{\text{INR}_2}{\text{SNR}_1})$ and sends at a power of $\frac{1}{\text{SNR}_2}(\frac{\text{INR}_2}{\text{SNR}_1}-1)$. The second receiver does joint decoding while the third receiver treats the message from the second transmitter as noise. This achieves the sum capacity within 2 bits.

*G.* $\text{INR}_2 \geq \text{SNR}_1$, $\text{SNR}_1 \geq \frac{\text{INR}_2}{1+\text{SNR}_2}$, $\text{SNR}_2 \geq \text{INR}_2$, $\text{INR}_3 \leq \text{SNR}_3$, $(\text{INR}_2+1)\text{INR}_3 \leq \text{SNR}_2$, $\text{SNR}_1(1+\frac{\text{SNR}_2}{1+\text{INR}_3}) \leq \text{INR}_2$

In this case, the second transmitter forms a single code-book of rate $\log(1+\frac{\text{SNR}_2}{1+\text{INR}_3})$ and sends at a power of $\frac{1}{1+\text{INR}_3}$. The second receiver does joint decoding while the third receiver treats the message from the second transmitter as noise. This achieves the sum capacity within 2 bits.

*H.* $\text{INR}_2 \geq \text{SNR}_1$, $\text{SNR}_1 \geq \frac{\text{INR}_2}{1+\text{SNR}_2}$, $\text{SNR}_2 \geq \text{INR}_2$, $\text{INR}_3 \geq \text{SNR}_3$, $(\text{INR}_2+1)\text{INR}_3 \leq \text{SNR}_2$

In this case, the second transmitter forms two codebooks and sends a superposition of these codebooks. The first codebook has rate $R_{2c} = \log(1+\frac{\text{INR}_3^2}{1+2\text{INR}_3+\text{SNR}_3(1+\text{INR}_3)})$ which is transmitted with a power of $\frac{\text{INR}_3}{1+\text{INR}_3}$. The second codebook has a rate of $R_{2p} = \log(1+\frac{\text{SNR}_2}{1+\text{INR}_3}) - \log(1+\text{SNR}_1)$ which is transmitted with a power of $\frac{1}{1+\text{INR}_3}$. The second receiver decodes these two messages and the message of the first transmitter jointly. The third receiver decodes $R_{2c}$ treating rest as noise, and then decodes the data of third user treating $R_{2p}$ as noise. This achieves sum capacity within 4 bits.

*I.* $\text{INR}_2 \geq \text{SNR}_1$, $\text{SNR}_1 \geq \frac{\text{INR}_2}{1+\text{SNR}_2}$, $\text{SNR}_2 \geq \text{INR}_2$, $\text{INR}_3 \geq \text{SNR}_3$, $(\text{INR}_2+1)\text{INR}_3 \geq \text{SNR}_2$, $\text{SNR}_1(1+\frac{\text{SNR}_2}{1+\text{INR}_3}) \geq \text{INR}_2$

In this case, the second transmitter forms two codebooks and sends a superposition of these codebooks. The first codebook has rate $R_{2c} = (\min(\log(1+\frac{\text{SNR}_2}{1+\text{INR}_2}), \log(1+\frac{\text{INR}_3^2}{1+2\text{INR}_3+\text{SNR}_3(1+\text{INR}_3)})) - 1)^+$ which is transmitted with a power of $\frac{\text{INR}_3}{1+\text{INR}_3}$. The second codebook has a rate of $R_{2p} = (\log(\frac{\text{INR}_2}{\text{SNR}_1}) - 1)^+$ and is transmitted at a power of $\frac{1}{1+\text{INR}_3}$. The third receiver decodes $R_{2c}$ treating rest as noise, and then decodes the data of third user treating $R_{2p}$ as noise. The second receiver decodes the three codebooks, two of the second transmitter and one of the first jointly. This achieves sum capacity within 4 bits.

*J.* $\text{INR}_2 \geq \text{SNR}_1$, $\text{SNR}_1 \geq \frac{\text{INR}_2}{1+\text{SNR}_2}$, $\text{SNR}_2 \geq \text{INR}_2$, $\text{INR}_3 \geq \text{SNR}_3$, $(\text{INR}_2+1)\text{INR}_3 \geq \text{SNR}_2$, $\text{SNR}_1(1+\frac{\text{SNR}_2}{1+\text{INR}_3}) \leq \text{INR}_2$, $\text{INR}_3 \leq \text{SNR}_2$

In this case, the second transmitter forms two codebooks and sends a superposition of these codebooks. The first codebook has rate $R_{2c} = (\min(\log(1+\text{INR}_3) - \log(1+\text{SNR}_1), \log(1+\frac{\text{INR}_3^2}{1+2\text{INR}_3+\text{SNR}_3(1+\text{INR}_3)})) - 1)^+$ which is transmitted with a power of $\frac{\text{INR}_3}{1+\text{INR}_3}$. The second codebook has a rate of $R_{2p} = \log(1+\frac{\text{SNR}_2}{1+\text{INR}_3})$ and is transmitted at a power of $\frac{1}{1+\text{INR}_3}$. The third receiver decodes $R_{2c}$ treating rest as noise, and then decodes the data of third user treating $R_{2p}$ as noise. The second receiver decodes the three codebooks, two of the second transmitter and one of the first jointly. This achieves sum capacity within 3 bits.

*K.* $\text{INR}_2 \geq \text{SNR}_1$, $\text{SNR}_1 \geq \frac{\text{INR}_2}{1+\text{SNR}_2}$, $\text{SNR}_2 \geq \text{INR}_2$, $\text{INR}_3 \geq \text{SNR}_3$, $(\text{INR}_2+1)\text{INR}_3 \geq \text{SNR}_2$, $\text{SNR}_1(1+\frac{\text{SNR}_2}{1+\text{INR}_3}) \leq \text{INR}_2$, $\text{INR}_3 \geq \text{SNR}_2$

In this case, the second transmitter forms a single codebook of rate $\log(1+\min(\text{SNR}_2, \frac{\text{INR}_3}{1+\text{SNR}_3}, \frac{\text{SNR}_2+\text{INR}_2-\text{SNR}_1}{1+\text{SNR}_1}))$ and sends at a power of 1. The second and the receiver does joint decoding. This achieves the sum capacity within 1 bit.

## References

[1] D. Blackwell, L. Breiman, and A. J. Thomasian, "The capacity of a class of channels," *Ann. Math. Stat.*, vol. 30, pp. 1229-1241, Dec. 1959.
[2] A. Raja, V. M. Prabhakaran, and P. Viswanath, "The Two User Gaussian Compound Interference Channel," *IEEE Transactions on Information Theory*, pp. 5100-5120, Nov 2009.
[3] L. Tassiulas and A. Ephremides, "Stability properties of constrained queueing systems and scheduling policies for maximum throughput in multihop radio networks," *IEEE Transactions on Automatic Control*, pp. 1936-1948, 1992.
[4] L. Meng, A. Zipf, and S. Winter, "Map-based mobile services: design, interaction, and usability." Springer; 1 edition, 2008.
[5] M. Kubale, "Graph Colorings". American Mathematical Society, 2004.
[6] V. Aggarwal, Y. Liu, and A. Sabharwal, "Message Passing in Distributed Wireless Networks," in Proc. *IEEE Internarional Symposium on Information Theory*, Jun-Jul 2009, Seoul, Korea.
[7] V. Aggarwal, Y. Liu, and A. Sabharwal, "Sum-capacity of interference channels with a local view: Impact of distributed decisions," submitted to *IEEE Transactions on Information Theory*, Oct 2009, available at arXiv:0910.3494v1.
[8] H. Sato, "The capacity of the Gaussian Interference channel using Strong interference", *IEEE Transactions on Information Theory*, vol. IT-27, pp. 786-788, Nov. 1981.
[9] V. Aggarwal, S. Avestimehr, and A. Sabharwal, "Distributed Universally Optimal Strategies for Interference Channels with Partial Message Passing," *in Proc. Allerton Conference on Communication, Control, and Computing,* Monticello, IL, Sept-Oct 2009.
[10] A. S. Avestimehr, S. N. Diggavi, and D. N. C. Tse, "Wireless network information flow: a deterministic approach," submitted to *IEEE Transactions on Information Theory*, Aug 2009, available at arXiv:0906.5394v2.
[11] G. Bresler, A. Parekh and D. Tse, "The approximate capacity of the many-to-one and one-to-many Gaussian interference channels," submitted to *IEEE Transactions on Information Theory*, Sept 2008, available at arXiv:0809.3554v1.
[12] B. Hajek and G. Sasaki, "Link scheduling in polynomial time," *IEEE Transactions on Information Theory* vol. 34(5), pp. 910-917, Sept 1988.